\documentclass[fleqn,usenatbib]{mnras}

\usepackage[T1]{fontenc}

\DeclareRobustCommand{\VAN}[3]{#2}
\let\VANthebibliography\thebibliography
\def\thebibliography{\DeclareRobustCommand{\VAN}[3]{##3}\VANthebibliography}

\usepackage{graphicx}	% Including figure files
\usepackage{amsmath}	% Advanced maths commands
\usepackage{amssymb}	% Extra maths symbols
\usepackage{threeparttable}
\usepackage{multirow}
\usepackage{soul,xcolor} % for text 's colour and strikethrough

\usepackage{newtxtext,newtxmath}
\title[A super-Eddington neutron star in NGC 2403]{NGC 2403 XMM4: evidence for a super-Eddington neutron star with a possible transient pulsation}

\author[W. Luangtip et al.]{
Wasutep Luangtip,$^{1}$$^{,2}$\thanks{E-mail: wasutep@g.swu.ac.th (WL)}
and Timothy P. Roberts$^{3}$
\\
$^{1}$Department of Physics, Faculty of Science, Srinakharinwirot University, Bangkok 10110, Thailand\\
$^{2}$National Astronomical Research Institute of Thailand, Chiang Mai 50180, Thailand\\
$^{3}$Centre for Extragalactic Astronomy and Department of Physics, Durham University, South Road, Durham DH1 3LE, UK
}

\date{Accepted XXX. Received YYY; in original form ZZZ}

\pubyear{2023}
\begin{document}
\label{firstpage}
\pagerange{\pageref{firstpage}--\pageref{lastpage}}
\maketitle

\begin{abstract}

We present a study of the X-ray source NGC 2403 XMM4 (4XMM J073702.2+653934) based on 20 years of archival observations with {\it XMM-Newton, Chandra, Swift} and {\it NuSTAR}.  Although it has previously been classified as an ultraluminous X-ray source (ULX), we show that its luminosity rarely, if ever, passes the $10^{39} \rm ~erg~s^{-1}$ threshold luminosity for a ULX.  It does, however, behave very similarly to ULXs, with its 0.3-10\,keV spectra well described by two thermal components, the softer of which behaves consistently with the expectations for an advection-dominated disc ($L\propto T^{1.49\pm0.85}$), and we find tentative evidence for an extra spectral component above 10 keV.  We also find moderately significant evidence for an absorption feature in one spectrum that could originate in an outflowing wind, although a cyclotron resonance scattering feature is also a possibility.  Most intriguingly, we find a possible transient pulsation at $\sim 3.32$ Hz in a short segment of one observation using an accelerated pulsation search.  This evidence suggests that NGC 2403 XMM4 is displaying many of the hallmarks of super-Eddington accretion at luminosities between $5 - 10 \times 10^{38} \rm ~erg~s^{-1}$ which, when considered alongside the putative pulsation, points to the presence of a neutron star as the accreting object this system.

\end{abstract}


%%%%%%%%%%%%%%%%%%%%%%%%%%%%%%%%%%%%%%%%%%%%%%%%%%

%%%%%%%%%%%%%%%%% BODY OF PAPER %%%%%%%%%%%%%%%%%%

\section{Introduction} \label{sec:Introduction}

Ultraluminous X-ray sources (ULXs) are defined as point-like X-ray sources with typical X-ray luminosities ($L_{\rm X}$) $>$ 10$^{39}$ erg s$^{-1}$ that are located in the extra-nuclear regions of galaxies.  Historically, the two main interpretations for such a relatively high luminosity have been either accretion at sub-Eddington rates onto a massive compact object such as an intermediate mass black hole (IMBH), or super-Eddington (or $\sim$Eddington) accretion onto a less massive object, e.g. a stellar mass black hole (BH) or even a neutron star (NS). ULXs are now becoming increasingly well studied, having been subject to dedicated observations for more than two decades (see e.g. \citealt{Kaaret2017,King2023} for recent reviews). One surprising result in the field was the first detection of a pulsating signal, from the ULX M82 X-2 \citep{Bachetti2014} which provided the first strong evidence that ULXs could be powered by super-Eddington NSs, i.e. NS-ULXs, in sharp contrast to the original assumption that they would most likely be powered by BHs.  Following this first detection of a pulsating NS-ULX -- hereafter PULX -- several more objects were revealed to have a pulsating signal in their observational data, with notable examples including NGC 5907 ULX-1 \citep{Israel2017a}, NGC 7793 P13 \citep{Furst2016,Israel2017b}, M51 ULX-7 \citep{Rodriguez2020} and NGC 1313 X-2 \citep{Sathyaprakash2019}; see also Table 2 of \citet{King2023} and references therein for other detected PULXs. In addition, NS-ULXs can also potentially be identified by the detection of cyclotron resonance scattering features (CRSF) imprinted on ULX spectra, due to the presence of strong magnetic fields ($\gtrsim 10^{12}$ G) in these objects (e.g., \citealt{Brightman2018,Walton2018b}).  Indeed, there is now the possibility that a majority (or, at least, a substantial fraction) of the whole ULX population could be powered by NSs, (c.f. \citealt{Walton2018a}) which implies by the observational definition of ULX luminosity that these NSs must be highly super-Eddington accretors. 

NGC 2403 is a nearby, Scd spiral galaxy located at distance of 4.2 Mpc \citep{Tully1988}, that provides a good opportunity for studying the brightest X-ray sources residing in a normal galaxy.  Indeed, the X-ray sources in this galaxy have been well studied with nearly 200 discrete sources having been identified \citep{Schlegel2003,Binder2015}. Among these, there is one source -- CXOU J073625.5+653540 -- that has been shown to be a persistent ULX and has been analysed by several studies using data from various observatories, e.g. \citet{Kotoku2000,Liu2005,Winter2006,Isobe2009}. Furthermore, the detection of a transient X-ray pulsar (CXOU J073709.1+653544) in this galaxy, with a maximum luminosity of 2.6 $\times$10$^{38}$ erg s$^{-1}$, has also been reported by \citet{Trudolyubov2007} in which it was argued that this luminosity exceeds the isotropic Eddington limit for a 1.4M$_{\odot}$ compact object.  Given that, there are at least two strong candidates for super-Eddington sources identified to date in this galaxy.

In fact, there is another X-ray source that has been classified as a ULX candidate in NGC 2403, in the catalogue published by \citet{Walton2011}: 2XMM~J073702.1+653935.\footnote{4XMM J073702.2+653934 in the latest 4XMM-DR13 catalogue.}  This is a relatively low luminosity ULX candidate, in the poorly-studied 'Eddington threshold' regime discussed in \cite{Earnshaw2017} ($L_{\rm X} $\la$ $10$^{39}$ erg s$^{-1}$).  It was previously studied in several papers, where it was referred to by a variety of different names: e.g., NGC 2403 X3 \citep{Liu2005}, NGC 2403 XMM4 \citep{Winter2006}, NGC 2403 source 5 \citep{Kotoku2000,Isobe2009} etc. (hereafter, we will refer to this source as NGC 2403 XMM4 for convenience). In fact, most works that studied NGC 2403 XMM4 were actually population studies and so the source was analysed together with, and then compared to, numerous other sources (see e.g. \citealt{Schlegel2003,Liu2005,Winter2006}). These previous studies found similar results: that the source could be well approximated by an absorbed, single component, hard power-law model (with photon index $\Gamma \sim$1.6 -- 1.9 ) or an absorbed thermal bremsstrahlung model with the plasma temperature of $\sim$5 keV, while its luminosity was reported to be $\sim$3 -- 5 $\times$10$^{38}$ erg s$^{-1}$. Given the source spectral shape and luminosity, it was proposed by \citet{Winter2006} that NGC 2403 XMM4 could be classified as a sub-Eddington intermediate mass BH.  However, in-depth analysis of this source is still required to firmly identify the class of compact object powering it.  

Given that NGC 2403 has been observed multiple times by various observatories since the beginning of this century, including several observations in the current decade, there is now sufficient archival data to provide us with the opportunity to analyse this source in detail. In this work, we provide a comprehensive analysis of the X-ray spectral and temporal characteristics of NGC 2403 XMM4, with the aim of investigating whether can we confirm that it is another super-Eddington source in NGC 2403, and if so whether we can determine the nature of its compact object.  The paper is laid out as follows. In Section~\ref{sec:Observations and data reduction}, we detail the observational data used in this study and explain how we produce the science products for analysis. The analysis methods and results are shown in Section~\ref{sec:Data analysis and Result} and these are then discussed in Section~\ref{sec:Discussion}, and our conclusions made in Section~\ref{sec:Conclusion}.

\section{Observations and data reduction} \label{sec:Observations and data reduction}

In order to study the properties of NGC 2403 XMM4, we gather observational X-ray data from the archives of the current generation of X-ray observatories -- i.e. {\it XMM-Newton},  {\it NuSTAR},  {\it Chandra} and  {\it Swift} -- that have sufficient data quality for timing and spectra analyses. The data used in this work are shown in Table~\ref{tab:obervations}, excepting the {\it Swift} data which are available in Appendix~\ref{sec:swift_obs}. The detail of observations analysed in this work and how they were reprocessed and converted into science products are described below.

%%%%%%%%%%%%%%%%%%%%%%%%%%%%%%%%%%%%%%%%%%%%%%%%%%

    \begin{table}
      \centering
      \caption{Observational data for NGC 2403 XMM4.}\label{tab:obervations}
      \smallskip
      \begin{threeparttable}
          \begin{tabular}{lccc}
          \hline
             Obs. ID & Obs. date &  Total GTI $^{a}$  & Count rate$^{b}$   \\
                          &                  &       (ks)          &          (counts s$^{-1}$)        \\
             \hline
           \multicolumn{4}{c}{\it XMM-Newton} \\  
         0164560901 & 2004-09-12 & 56.2/62.1/63.3 & 0.062 \\
	0870870201~[1]$^{\dagger}$ & 2020-10-25 & 8.1/11.9/13.2 &  0.062 \\
	0870870401~[2]$^{\dagger}$ & 2021-04-26 & 21.2/24.7/24.5 & 0.073 \\
	0870870501~[2]$^{\dagger}$ & 2021-04-28 & 20.7/23.6/23.3 & 0.073 \\
	0870870801 & 2021-04-22 & 21.4/23.6/23.5 & 0.042 \\
 		\hline
           \multicolumn{4}{c}{\it NuSTAR} \\  
         50610001002~[1]$^{\dagger}$ & 2020-10-27 & 104.8/104.9 & 0.001 \\
	50610001004~[2]$^{\dagger}$ & 2021-04-27 & 103.4/103.4 & 0.003 \\
	50610002002~[2]$^{\dagger}$ & 2021-04-29 & 106.0/106.0 & 0.002 \\
		\hline
           \multicolumn{4}{c}{\it Chandra} \\  
           2014 & 2001-04-17 & 35.2 & 0.046 \\
           4628 &  2004-08-23 & 47.4 & 0.033 \\
           4629 &  2004-10-03 & 45.5 & 0.032 \\
           4630 &  2004-12-22 & 50.8 & 0.030 \\
             \hline
         \end{tabular}
         \begin{tablenotes}
         \item \textit{Note.} $^{a}$The total (summed) value of good time intervals (GTI) for each instrument; for {\it XMM-Newton}, it is the total GTI of pn, MOS1 and MOS2 detectors, respectively, while for {\it NuSTAR}, it is that of FPMA and FPMB detectors, respectively. $^{b}$The background-subtracted, instrument count rate of the source; the value is obtained from pn and ACIS-S detectors in the 0.3 -- 10 keV band for {\it XMM-Newton} and {\it Chandra}, respectively, while for {\it NuSTAR}, the 3.0-30\,keV count rate obtained from the FPMA detector is reported. $^{\dagger}$The number in the square bracket indicates the epoch of data used in the broadband spectral analysis of {\it XMM-Newton} + {\it NuSTAR} data (see Section~\ref{sec:Broad band spectral model}). 
         \end{tablenotes}
      \end{threeparttable}
    \end{table}

%%%%%%%%%%%%%%%%%%%%%%%%%%%%%%%%%%%%%%%%%%%%%%%%%%

\subsection{{\it XMM-Newton} observations}

We collected all {\it XMM-Newton} archival observations that contain data from the X-ray source NGC 2403 XMM4; the data were downloaded from the {\it XMM-Newton} Science Archive (XSA).\footnote{\url{https://www.cosmos.esa.int/web/xmm-newton/xsa}}  However, the {\it XMM-Newton} observations analysed in this work are ones in which the summation of good exposure times (defined as the good time intervals, or GTIs) for each detector after the removal of background flaring periods is $\ga$ 10 ks, to ensure that we obtained sufficient data quality for our analysis. Using this criterion, we ended up with the five {\it XMM-Newton} observations presented in Table~\ref{tab:obervations}. However, we note that there are another five observations in which the total GTIs are insufficient for analysing in depth (i.e., Obs ID 0150651101, 0150651201, 0301860301, 0729560901 and 0870870301); here we were, at least, able to measure and report their individual fluxes, as shown in Fig.~\ref{fig:lightcurve}, using a similar method to that we used to measure the flux of individual {\it Swift} observations (see Section~\ref{sec:Spectral analysis}).

The observational data were reprocessed following the standard procedure suggested on the {\it XMM-Newton} data analysis webpage\footnote{\url{https://www.cosmos.esa.int/web/xmm-newton/sas-threads}} using the Science Analysis System ({\sc sas}) software version 21.0.0 with the latest calibration files at the time that the data were reprocessed (May 2023). In detail, we reprocessed the pn and MOS data using the SAS packages {\sc epproc} and {\sc emproc}, respectively. The reprocessed files were then inspected for any periods that were affected by background flaring events by visually checking the high energy light curves in the 10-12\,keV energy band.  If there were any observing periods in which the count rates exceeded $\ga$0.4 count s$^{-1}$ for the pn detector or $\ga$0.35 count s$^{-1}$ for MOS1 or MOS2, then the data from these periods were removed from the reprocessed event files. For spectral analysis,  we created the spectral files and their corresponding redistribution matrix file (RMFs) and auxiliary response files (ARFs) for all observations using the default event flags and patterns for each detector, using a circular data extraction region with radius $25\arcsec$ located at the source position.  We used a source-free region of $32\arcsec$ radius located nearby and on the same detector chip as the source to extract a background spectrum. We note that for the {\it XMM-Newton} spectra that we analysed  simultaneously with {\it NuSTAR} data, we also applied a correction to the effective areas in the ARFs by setting the parameter {\sc applyabsfluxcorr} in the SAS task {\sc arfgen} to `yes'. This could help to avoid a flux discrepancy between {\it XMM-Newton} data and external flux data (i.e., those of {\it NuSTAR}).\footnote{\url{https://xmmweb.esac.esa.int/docs/documents/CAL-TN-0230-1-3.pdf}}  Finally, all spectra were grouped by the SAS task {\sc specgroup} setting the parameters mincounts $=$ 20 and oversample $=$ 3, to utilise the chi-squared statistic in spectral modelling.

We also prepared files for timing analysis, with the goal of searching for pulsations.  We first performed a barycentric correction to the cleaned event files using the SAS command {\sc barycen}, and then created both pn and MOS light curves for all observations using the same source and background regions as used in the spectral extraction,  and all science events with the appropriate flags. The binning of the resulting light curves was set to be equal to the temporal resolution of the science mode of each instrument used during the observation, i.e., 73.4 ms for the pn and 2.6 s for MOS1 and MOS2.  We also extracted the light curves in three separate energy bands: 0.3 -- 2 keV (soft band), 2 -- 10 keV (hard band) and 0.3 -- 10 keV (full band).

\subsection{{\it NuSTAR} observations}

The three {\it NuSTAR} archival observations were reprocessed using  the task {\sc nupipeline} of {\it NuSTAR} Data Analysis Software ({\sc nustardas}) version 2.1.2, which is part of the {\sc heasoft} package.\footnote{\url{https://heasarc.gsfc.nasa.gov/docs/software/heasoft/}} We also applied a barycentric correction to the cleaned event files using the {\sc ftool} task {\sc barycorr}.\footnote{\url{https://heasarc.gsfc.nasa.gov/ftools/caldb/help/barycorr.html}} Then, spectra and timing products were extracted from the cleaned event files using the task {\sc nuproducts}; the source region for extracting science products was defined as a circular region with a radius of 40$\arcsec$ located at the source position, while the background was taken from a circular, source-free region with a radius of $\sim$120$\arcsec$ -- 150$\arcsec$ close to the source region. Similar to {\it XMM-Newton} spectra, the {\it NuSTAR} spectra were then grouped to have a minimum of 20 counts per bin for spectral analysis.

\subsection{{\it Chandra} observations}

There are four {\it Chandra} observations in which NGC 2403 XMM4 lies inside the field of view, three of which were taken in 2004 and are ToO observations of supernova 2004dj in NGC 2403 \citep{Pooley2004}. All {\it Chandra} data were download from the {\it Chandra} data archive\footnote{\url{https://cda.harvard.edu/chaser/}} and then reprocessed using the standard script {\sc chandra\_repro} which is part of the Chandra Interactive Analysis of Observations ({\sc ciao}) software version 4.15.\footnote{\url{https://cxc.cfa.harvard.edu/ciao/}} Data were checked for background flaring, which was not significant in 3/4 observations, and only minimally affected the fourth (Obs. ID 2014, from which $\la 2$ per cent of data was removed using the task {\sc lc\_sigma\_clip}. In addition, we also applied a barycentric correction to the time column in the cleaned event files using the {\sc ciao} tool {\sc axbary}.

We then extracted science products for all observations in the full (0.3-10\,keV) band using the {\sc ciao} task {\sc srcflux}; this task produced both spectral and timing products (i.e., light curves) and provided these for each individual dataset as well as for merged data. Here, the source data were extracted from a circle of 5$\arcsec$ radius, while the background region was a source-free annulus of 7$\arcsec$ inner radius and 12$\arcsec$ outer radius; both regions were centred on the source location. Note that for the light curves, the time resolution that we used for binning the data is set equal to the nominal frame time of the {\it Chandra} ACIS detectors in the observations used in this work, i.e. $\sim$3.2 seconds.

\subsection{{\it Swift} observations}

The {\it Swift} data for NGC 2403 XMM4 were selected from observations in the {\it Swift} data archive in which the ULX is within the field of view. In total, we obtained 27 useful observations that have sufficient net photon counts to constrain the ULX count rate and flux; the observational details are summarised and shown in Table~\ref{tab:swift_obs} in Appendix~\ref{sec:swift_obs}.

The {\it Swift} data were reprocessed to create level 2 cleaned event-list files using the pipeline {\sc xrtpipeline} version 0.13.7, which is included in the {\sc heasoft} software package. Then, the spectrum of each observation was extracted from a circular aperture with radius of 35$\arcsec$ centred on the ULX position.  A background spectrum was obtained from a source-free annulus around the source position, which has 50$\arcsec$ inner and 100$\arcsec$ outer radii. The detail of the {\it Swift} spectral analysis will be described in Section~\ref{sec:Spectral analysis}.

\section{Data analysis and Results} \label{sec:Data analysis and Result}

In this section we present the results from spectral and temporal analyses of the data from NGC 2043 XMM4.  As our investigation is largely focussed on whether this object is a super-Eddington accretor, given its previous identification as a ULX candidate, we will compare its properties to those of ULXs throughout this section.

\subsection{Spectral analysis} \label{sec:Spectral analysis}

Throughout this paper, spectral analysis was performed using the X-ray spectral fitting package {\sc xspec}, version 12.13.0 \citep{Arnaud1996}.\footnote{\url{https://heasarc.gsfc.nasa.gov/xanadu/xspec/}} We also assume that the observed X-ray data are absorbed by material along the line of sight in the direction to NGC 2403 XMM4 so that two absorption components ({\sc tbabs$\times$\sc tbabs} in {\sc xspec}) were added to every spectral model we used in this paper.  The first component represents the absorption by neutral material in our own Galaxy, for which the value was frozen at $0.04\times10^{22}$ atoms cm$^{-2}$ \citep{Dickey1990},\footnote{Calculated using the HEASARC N$_{\rm H}$ web tool: \url{https://heasarc.gsfc.nasa.gov/cgi-bin/Tools/w3nh/w3nh.pl}}, while the other component can be interpreted as the absorption by material in the host galaxy and/or by material in the close proximity of the source itself, and we refer to this as the external absorption column density hereafter.  If not otherwise specified, we allowed the value of the latter component to be a free parameter during the spectral fitting, and the spectra were normally fitted in a range of 0.3 -- 10 keV.  The source luminosities were calculated assuming a distance of 4.2 Mpc \citep{Tully1988}. In addition, for those observatories where multiple detectors take simultaneous data, a constant component was added to the models used to fit the data to allow for any small differences between detector responses.  The errors on spectral fits are quoted as 90 per cent confidence ranges of parameters unless otherwise specified.

\subsubsection{Single component models}

\begin{figure*}
 \includegraphics[width=\textwidth]{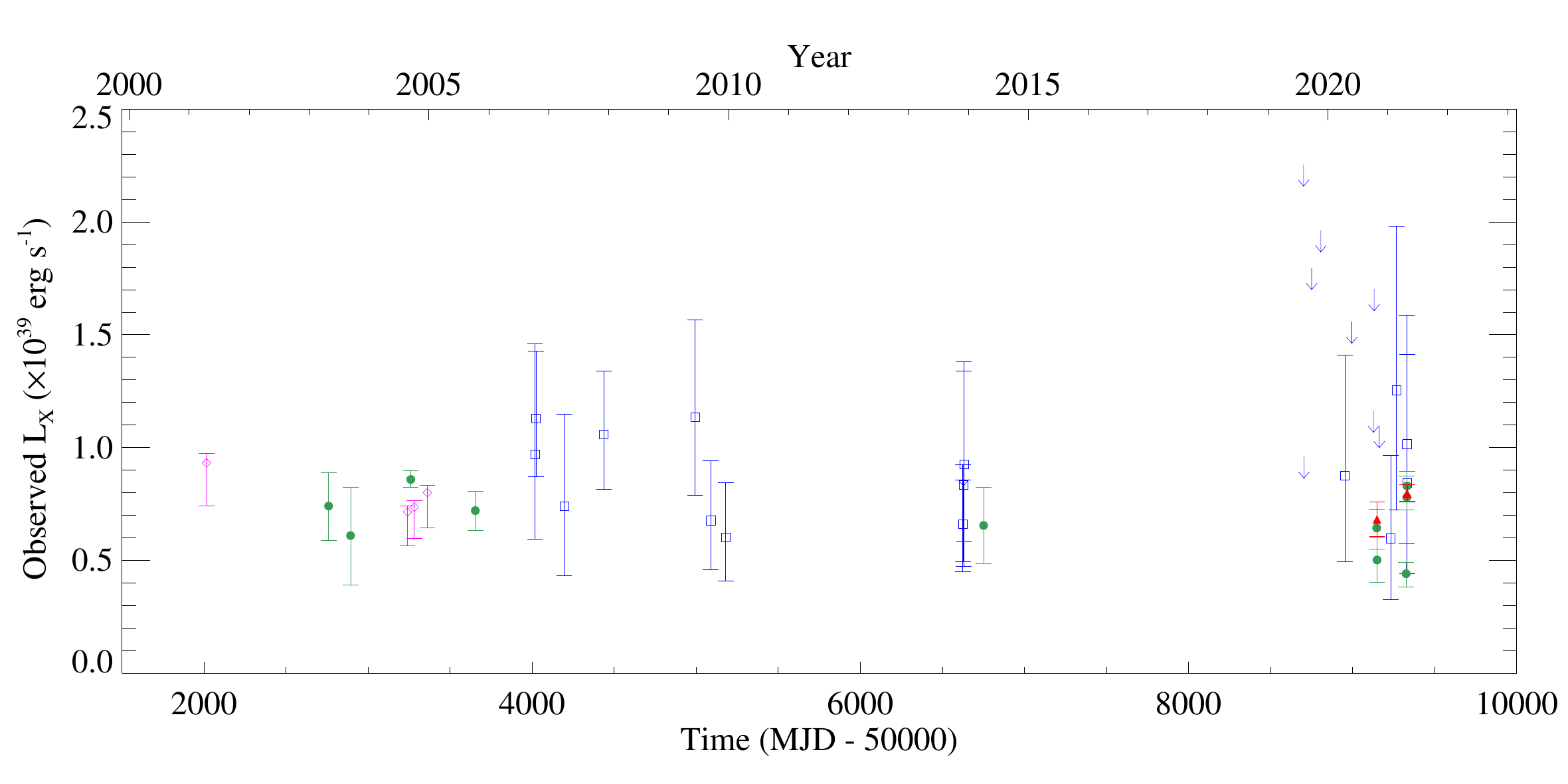}
 \caption{Twenty year light curve of NGC 2403 XMM4 obtained from {\it Swift} (blue boxes), {\it Chandra} (magenta diamonds), {\it XMM-Newton} (green circles) and {\it NuSTAR} (red triangles) observations.  The observed luminosity is reported in the 0.3-10\,keV energy band, with {\it Chandra} and {\it NuSTAR } values extrapolated to that band from their best-fitting parameters.}
 \label{fig:lightcurve}
\end{figure*}

We began the analysis by fitting the spectra of individual {\it Swift} observations. However, due to the low number of net photons of the source ($\sim$5 -- 50 counts per observation), the {\sc xspec} statistic {\it cstat} was applied to fit the {\it Swift} spectra, that was kept unbinned for this analysis.  We also simplified the spectral fitting by using a single component absorbed power-law model and reduced the number of free parameters by freezing the absorption column density and power-law photon index at  $0.2\times10^{22}$ atoms cm$^{-2}$ and 1.6, respectively; note that the fixed values of these two parameters were obtained from the best-fit values of the merged {\it Swift} spectrum fitted by the same model (see Table~\ref{tab:swift_fit}). Moreover, these parameter values were also consistent with those we could obtain from individual {\it XMM-Newton} spectra fitted by an absorbed power-law model; this implies that the parameter values are a good (albeit rough) representation of the power-law-like spectral shape of this source.  Given that the two parameters are frozen, the only free parameter in the fitting is the power-law normalisation which was calculated in terms of flux using the {\sc xspec cflux} convolution component and then converted into luminosity.  Fig.~\ref{fig:lightcurve} shows the long term luminosity of NGC2403 XMM4 as observed by {\it Swift } in the interval 2006 -- 2021, and augmented by single epoch flux measurements from the other detectors considered.

%%%%%%%%%%%%%%%%%%%%%%%%%%%%%%%%%%%%%%%%%%%%%%%%%%
%%%%%%%%%%%%%%%%%%%%%%%%%%%%%%%%%%%%%%%%%%%%%%%%%%

To investigate the {\it Swift} spectra further, we maximised the spectral quality by merging all the individual spectra to become a single averaged spectrum using the {\sc ftools} task {\sc addspec}\footnote{\url{https://heasarc.gsfc.nasa.gov/ftools/caldb/help/addspec.txt}} and grouped the data to have a minimum of 20 counts per bin. We then fitted the spectrum with two traditional, single component models widely used for ULXs, an absorbed power-law and an absorbed multi-colour disc blackbody (MCD; {\sc diskbb} in {\sc xspec}). The result is shown in Table~\ref{tab:swift_fit} in which both models are seen to fit to the data equally well, while the best-fitting values of the MCD temperature and power-law photon index are consistent with past studies of ULXs (e.g. \citealt{Swartz2004,Gladstone2009}).

Another point we would emphasise is that the averaged luminosity of NGC 2403 XMM4 obtained from {\it Swift} data is $\sim$0.6 -- 0.8 $\times$ 10$^{39}$ erg s$^{-1}$, which lies below the traditional threshold of ULX luminosity, although individual data points show that the source occasionally rises to luminosities above 10$^{39}$ erg s$^{-1}$.  Given this behaviour, we discuss whether the source should be regarded as a ULX in Section~\ref{is it a ULX}.

%%%%%%%%%%%%%%%%%%%%%%%%%%%%%%%%%%%%%%%%%%%%%%%%%%

    \begin{table*}
      \centering
      \caption{The best fitting results from the merged {\it Swift} spectrum, using single component models}\label{tab:swift_fit}
      \smallskip
      \begin{threeparttable}
          \begin{tabular}{lcccc}
          \hline
             Model & $N_{\rm H}$$^{a}$ &  $kT_{\rm in}$ / $\Gamma$$^{b}$  & $\chi^{2}$/d.o.f.$^{c}$   & $L_{\rm X}$$^{d}$ \\
             & ($\times$ 10$^{22}$ cm$^{-2}$)& (keV)  &  &($\times$ 10$^{39}$ erg s$^{-1}$)  \\
             \hline
             
MCD & $<0.23$ & $1.52_{-0.27}^{+0.39}$ & 9.85/17 & $0.63_{-0.40}^{+0.05}$ \\ 
Power-law & $0.26_{-0.13}^{+0.17}$ & $1.64_{-0.26}^{+0.28}$ & 10.2/17 & $0.78\pm0.13$ \\ 

             \hline
         \end{tabular}
         \begin{tablenotes}
         \item \textit{Notes.} $^{a}$External absorption column density.  $^{b}$Inner disc temperature or photon index for MCD and power-law models, respectively. $^{c}$$\chi^{2}$ and number of degrees of freedom for the best fit. $^{d}$Observed X-ray luminosity in the 0.3-10\,keV energy band.
         \end{tablenotes}
      \end{threeparttable}
    \end{table*}

%%%%%%%%%%%%%%%%%%%%%%%%%%%%%%%%%%%%%%%%%%%%%%%%%%

Next we analysed the {\it Chandra} data.  Given these observations were taken in the period 2001-2004 we note this extends coverage of NGC 2043 XMM4 to earlier times than the {\it Swift} data.  We began by analysing each individual {\it Chandra} dataset in the 0.5 -- 7 keV energy band, with the spectra binned to have a minimum of 15 counts per bin, and used the same absorbed, single component models as per the {\it Swift} merged data.  We found that although both models could yield acceptable fitting results, the absorbed MCD model is able to describe the individual {\it Chandra} spectra better (a total best-fitting $\chi^{2}$ of 317 for the MCD model versus 344 for the power-law, for a total of 309 degrees of freedom).  The calculated fluxes from the MCD model are shown in Fig.~\ref{fig:lightcurve} which indicates that the flux obtained from {\it Chandra} data is roughly constant.  Furthermore, the best-fitting MCD temperatures of all the spectra are consistent, with their values $\approx$1.9$\pm$0.2 keV, which could imply that they are all in a similar spectral state.  Interestingly, the temperatures for the disc component are relatively high when compared to stellar-mass Galactic BHs in the thermal dominant state (c.f. fig. 1 of \citealt{Miller2004}; see also \citealt{Done2007} and references therein), however similar high disc temperatures can sometimes be seen for Z-type NS low-mass X-ray binaries (e.g., \citealt{Mondal2018,Homan2018}).  Similar high disc temperatures are also seen in the broadened disc-like regime (\citealt{Sutton2013}, see also \citealt{Luangtip2016}), that is sometimes observed for ULXs.  

Given the obvious similarity of the individual {\it Chandra} spectra, we decided to improve the spectral data quality by merging all the spectra together.  The improvement in signal to noise was significant enough that the data was appropriate for fitting with more complicated double component models, which we detail in the next sub-section.

\subsubsection{Double component model} \label{sec:Double component model}

\begin{figure*}
 \includegraphics[width=8.8cm]{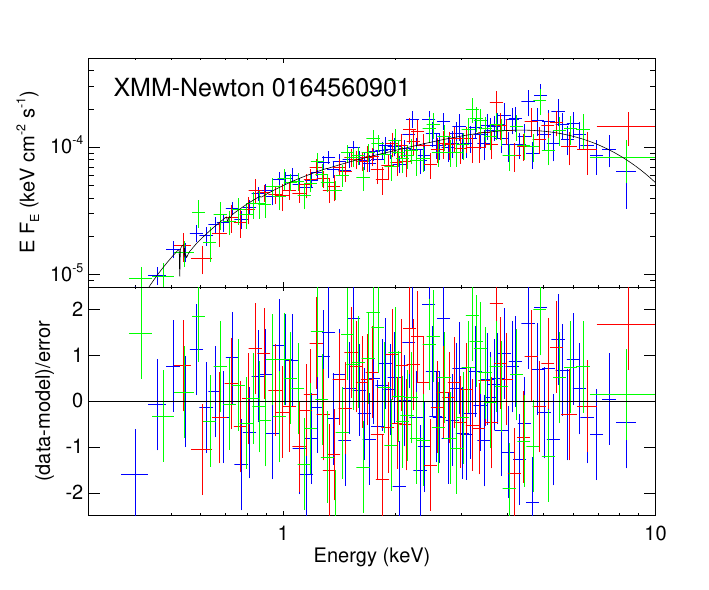} ~\includegraphics[width=8.8cm]{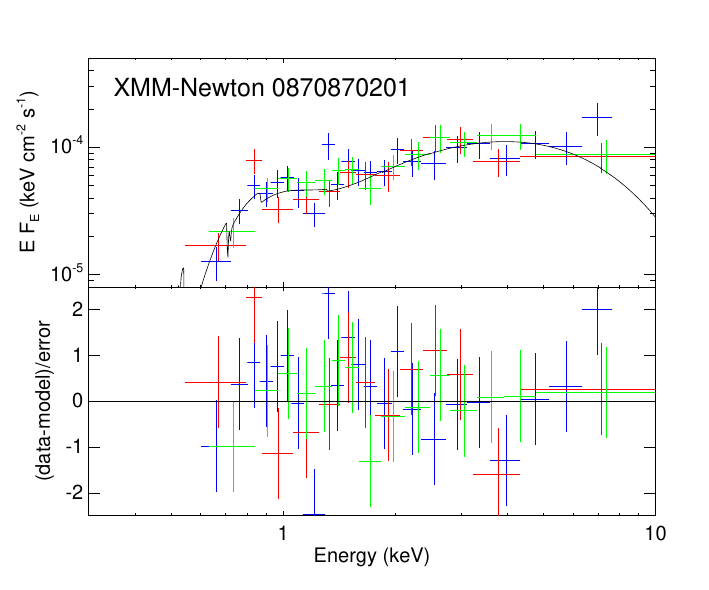}
 \includegraphics[width=8.8cm]{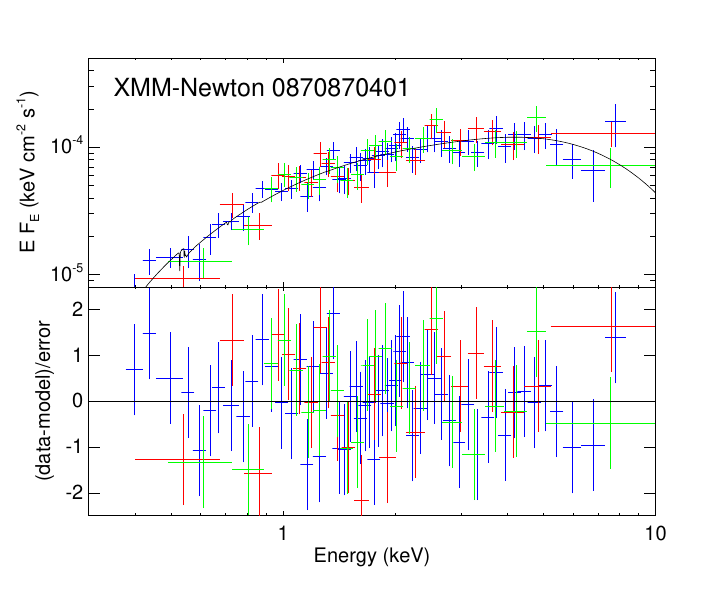} ~\includegraphics[width=8.8cm]{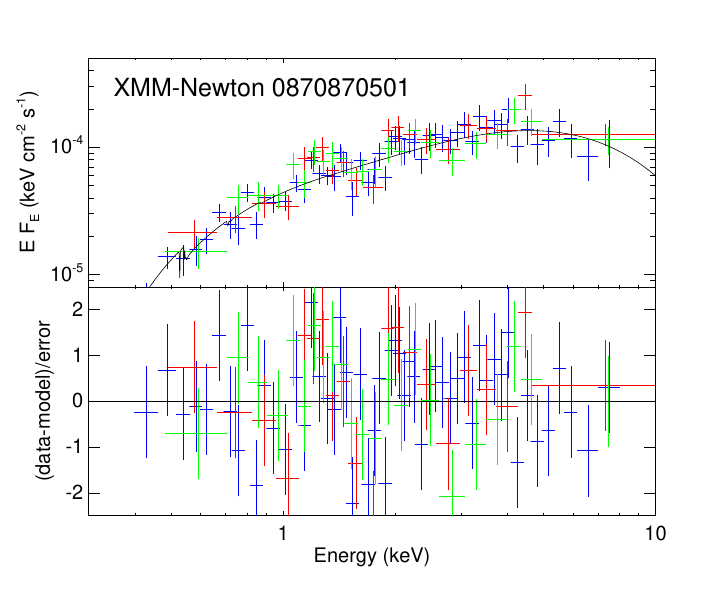}
 \includegraphics[width=8.8cm]{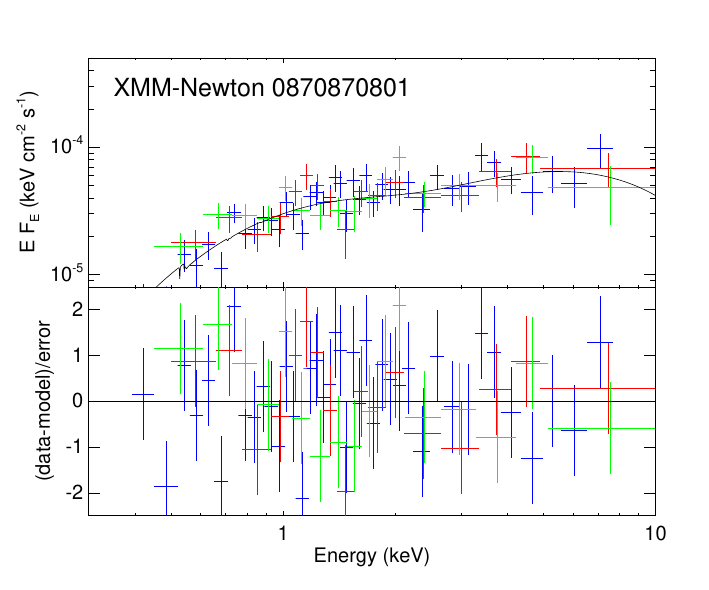} ~\includegraphics[width=8.8cm]{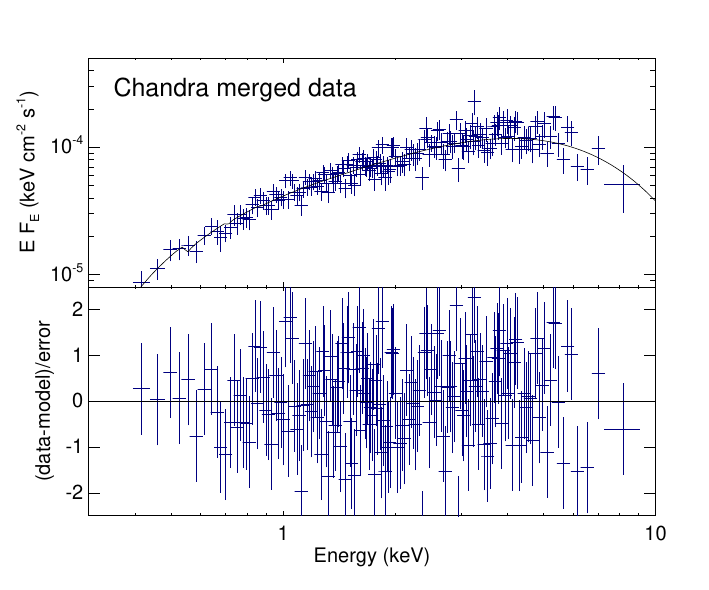}
 \caption{Individual {\it XMM-Newton} and merged {\it Chandra} spectra of NGC 2403 XMM4 with the best-fitting two component model in each case.  Top panels: best fitting model (solid-black line; see Section~\ref{sec:Double component model}) overlaying the data.  The pn, MOS1 and MOS2 data are plotted in blue, red and green, respectively; the dark blue in the bottom-right plot represents the merged {\it Chandra} spectrum. Bottom panels: the fitting residuals showing the difference between the model and the data in multiples of $\sigma$, where this is the $1\sigma$ uncertainty in the value of each data point.}
 \label{fig:xmm_spectra}
\end{figure*}

The individual {\it XMM-Newton} spectra and merged {\it Chandra} spectrum have sufficient data quality to enable the fitting of more complex models than the single component fits in the previous section.  Here, we attempt two component fits based on two thermal components, as is now widely accepted in ULX spectroscopy \citep{Stobbart2006,Koliopanos2017,Walton2020}, although the physical interpretation of these components varies.  We choose to model the data using a blackbody continuum and a MCD -- {\sc tbabs}$\times${\sc tbabs}({\sc bbody} + {\sc diskbb})  in  {\sc xspec } -- where, for example, the soft component may be the optically thick base of an outflowing wind (as detected by e.g. \citealt{Pinto2016}), and the harder component may come from the innermost regions of the accretion flow.  The fitting results are shown in Fig.~\ref{fig:xmm_spectra} and best-fitting values are shown in the top panel of Table~\ref{tab:two_component_fitting}; the calculated best-fit luminosities are also displayed in Fig.~\ref{fig:lightcurve} to further enhance the long timescale light curve of NGC 2403 XMM4.  The model provides an excellent fit to the data, with reduced $\chi^2$ ($\chi^2_{\nu} = \chi^2 /$dof) values below unity in 5 out of 6 cases. However, the spectrum of {\it XMM-Newton} observation 0870870501 has a slightly higher, but statistically acceptable, value of $\chi^2_{\nu} =1.13$, corresponding to a p-value for the null hypothesis probability of 0.2. In fact, the spectral data from this observation appear to exhibit structured fitting residuals at $\sim$1 keV and $\sim$1.7 keV (see Fig.~\ref{fig:xmm_spectra} middle-right image); we will return to this point in section~\ref{sec:spectral residuals}.  

\begin{figure*}
 \includegraphics[width=\columnwidth]{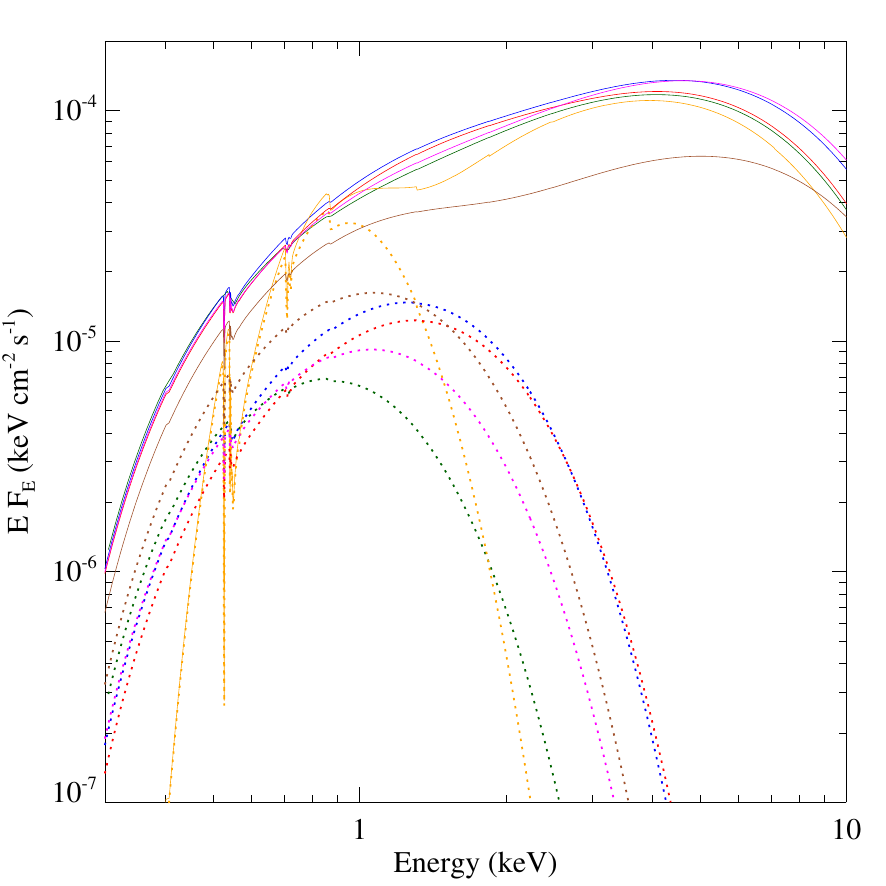}~\includegraphics[width=\columnwidth]{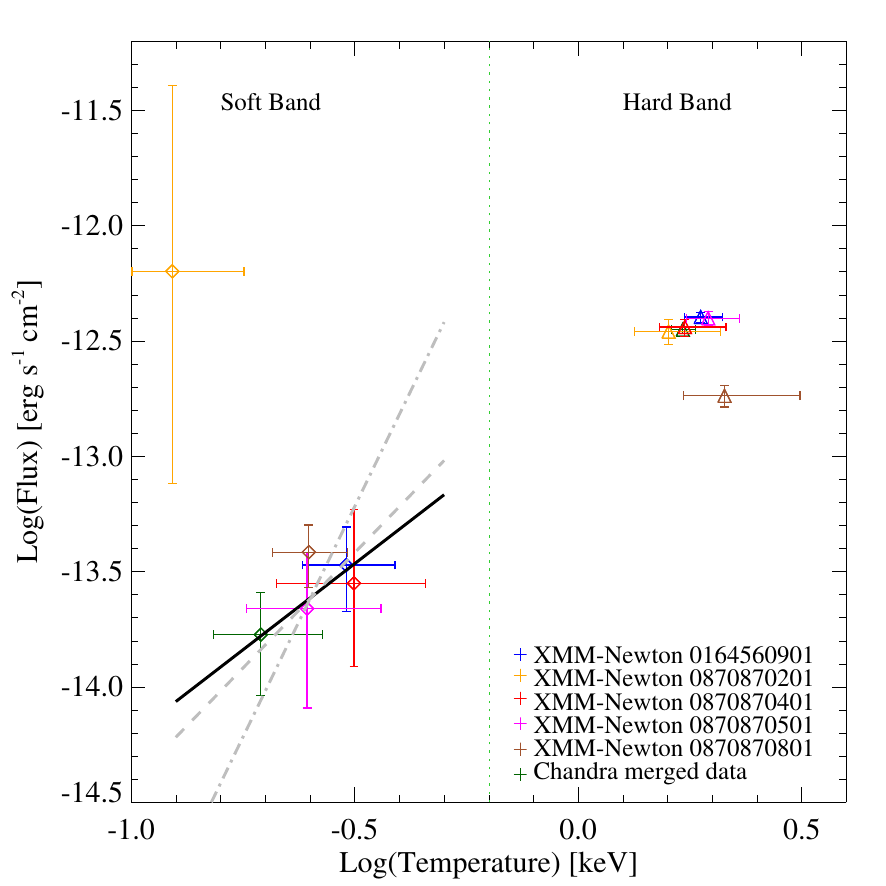}
 \caption{Left panel: the best-fitting two component models of individual {\it XMM-Newton} and merged {\it Chandra} spectra (solid line) along with the contribution from the BB component (dashed line), showing the trend of soft component evolution with the luminosity. Right panel: the relation between the temperature and the intrinsic flux of both individual model components; the data points towards the lower left plotted with a diamond symbol are those of the BB component, while those in the upper right plotted with triangle are the MCD component. The solid black line shows the best-fitting, linear correlation between logarithms of flux and temperature of the BB component of $L$ $\propto$ $T^{1.49}$, while the grey dashed and grey dashed-dot lines outline the $L$ $\propto$ $T^{2}$ and $L$ $\propto$ $T^{4}$ correlations, respectively. Note that the plotted data are obtained from fitting with the $N_{\rm H}$ tied, excepting for that of {\it XMM-Newton} 0870870201 in which the plotted data are adopted from the individual fit (see text for detail.)}
 
 \label{fig:L-T_correlation}
\end{figure*}

In addition, recent work on several ULXs (e.g. \citealt{Robba2021,Barra2022}) has shown correlations between the component fluxes and temperatures in similar two-component models. In fact, starting from \cite{Miller2013} and the following work (e.g. \citealt{Middleton2015,Walton2017}), the correlations, especially in the case of the soft component, tend to be detected when ULXs do not show strong variability in their absorption column. Here, we also check if this is the case for this source.  Excepting for the dataset 0870870201, the best-fit column density values are consistent within errors.  So, excluding that observation, we checked if any correlation is present by repeating our spectral fits with the column density held fixed across the five remaining datasets. A global $N_{\rm H}$ of 0.05 $\pm$ 0.02 $\times$ $10^{22}$~atom~cm$^{-2}$ was obtained (see lower part of Table~\ref{tab:two_component_fitting}), while the best-fit spectra and the relation between the temperature and the intrinsic flux of both soft and hard components are plotted in Fig.~\ref{fig:L-T_correlation}. Indeed, some degree of positive correlation between the soft component temperature and the flux is suggested in the plots although a Pearson's correlation ($r$) measurement shows that the correlation is not statistically significant. However, this could be driven by a second outlier dataset, 0870870801, with a significantly lower flux in its hard component, and a best-fit spectral shape suggestive of two distinct components, similar in form to the hard ultraluminous regime as defined by \citet{Sutton2013}.  Interestingly, if we also remove the outlier dataset from the correlation measurement, a linear correlation for the remaining four datasets is detected using a Pearson correlation ($r$ = 0.94), with a best-fit slope of 1.49 $\pm$ 0.85 ($1\sigma$ error), implying a luminosity ($L$) - temperature ($T$) correlation for the BB component in which $L$ $\propto$ $T^{1.49\pm0.85}$.

%%%%%%%%%%%%%%%%%%%%%%%%%%%%%%%%%%%%%%%%%%%%%%%%%%
%%%%%%%%%%%%%%%%%%%%%%%%%%%%%%%%%%%%%%%%%%%%%%%%%%

%%%%%%%%%%%%%%%%%%%%%%%%%%%%%%%%%%%%%%%%%%%%%%%%%%

    \begin{table*}
      \centering
      \caption{Best fits to the individual {\it XMM-Newton} and merged {\it Chandra} spectra, as modelled by a BB plus MCD model}\label{tab:two_component_fitting}
      \smallskip
      \begin{threeparttable}
          \begin{tabular}{lccccc}
          \hline
             Spectrum & $N_{\rm H}$ & $kT^{a}$ & $kT_{\rm in}$ & $\chi^{2}$/d.o.f.   & $L_{\rm X}$ \\
             & ($\times$ 10$^{22}$ cm$^{-2}$)& (keV) &(keV)  &  &($\times$ 10$^{39}$ erg s$^{-1}$)  \\
             \hline
 \multicolumn{6}{c}{Individual $N_{\rm H}$} \\             
0164560901 & $0.06_{-0.04}^{+0.06}$ & $0.27_{-0.08}^{+0.10}$ & $1.84_{-0.15}^{+0.21}$ & 187.56/203 & $0.86\pm0.04$ \\ 
0870870201 & $0.67_{-0.45}^{+0.42}$ & $0.12_{-0.03}^{+0.06}$ & $1.59_{-0.28}^{+0.42}$ & 44.01/44 & $0.64_{-0.09}^{+0.08}$ \\ 
0870870401 & $0.03_{-0.03}^{+0.10}$ & $0.37_{-0.18}^{+0.16}$ & $1.80_{-0.34}^{+0.52}$ & 83.42/97 & $0.78_{-0.06}^{+0.09}$ \\ 
0870870501 & $0.06_{-0.06}^{+0.13}$ & $0.23_{-0.08}^{+0.17}$ & $1.93_{-0.22}^{+0.32}$ & 102.19/91 & $0.83_{-0.07}^{+0.06}$ \\ 
0870870801 & $<0.12$ & $0.29_{-0.09}^{+0.05}$ & $2.33_{-0.61}^{+0.79}$ & 71.07/68 & $0.44_{-0.06}^{+0.05}$ \\ 
Merged Chandra & $0.03_{-0.03}^{+0.08}$ & $0.21_{-0.07}^{+0.12}$ & $1.72_{-0.10}^{+0.13}$ & 168.52/183 & $0.73\pm0.03$ \\ 
 
  	 \hline
 \multicolumn{6}{c}{$N_{\rm H}$ tied together} \\  
0164560901 & \multirow{5}{*}{$0.05_{-0.02}^{+0.03}$} & $0.30_{-0.07}^{+0.08}$ & $1.88_{-0.16}^{+0.21}$ & 189.86/204 & $0.86_{-0.03}^{+0.04}$ \\ 
0870870401 & & $0.32_{-0.13}^{+0.12}$ & $1.73_{-0.23}^{+0.37}$ & 84.30/98 & $0.77_{-0.05}^{+0.07}$ \\ 
0870870501 & & $0.25_{-0.08}^{+0.09}$ & $1.95_{-0.22}^{+0.31}$ & 102.82/92 & $0.83_{-0.05}^{+0.06}$ \\ 
0870870801 & & $0.25_{-0.05}^{+0.05}$ & $2.12_{-0.45}^{+0.83}$ & 72.22/69 & $0.43_{-0.04}^{+0.05}$ \\ 
Merged Chandra & & $0.20_{-0.05}^{+0.06}$ & $1.71_{-0.10}^{+0.12}$ & 168.76/184 & $0.73_{-0.03}^{+0.03}$ \\        
             
          \hline
         \end{tabular}
         \begin{tablenotes}
         \item \textit{Notes.} $^{a}$The temperature of the BB component. The remaining column definitions are as per Table~\ref{tab:swift_fit}. Note that as we do not included observation 0870870201 in the joint fitting, it is not in the lower half of the table. 
         \end{tablenotes}
      \end{threeparttable}
    \end{table*}

%%%%%%%%%%%%%%%%%%%%%%%%%%%%%%%%%%%%%%%%%%%%%%%%%%

%%%%%%%%%%%%%%%%%%%%%%%%%%%%%%%%%%%%%%%%%%%%%%%%%%
%%%%%%%%%%%%%%%%%%%%%%%%%%%%%%%%%%%%%%%%%%%%%%%%%%

\subsubsection{Narrow absorption-like residuals} \label{sec:spectral residuals}

\begin{figure}
\centering
 \includegraphics[width=9.2cm,height=8.2cm]{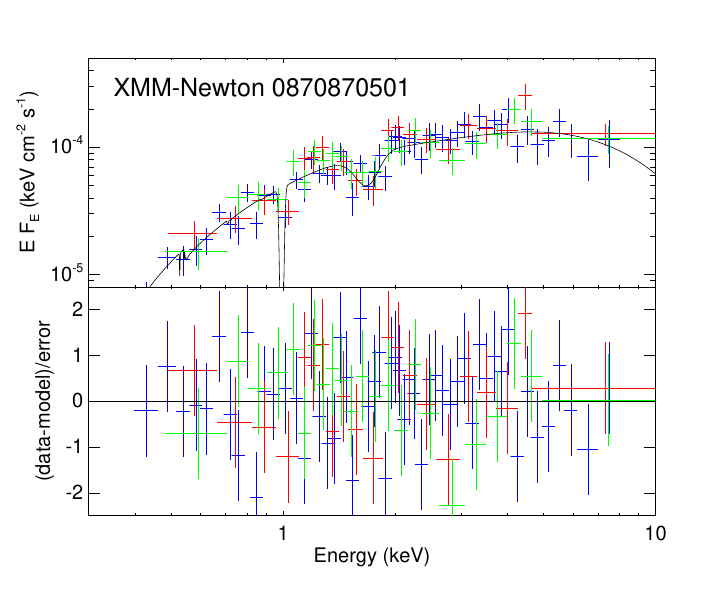}~ %\includegraphics[width=9.2cm,height=8cm]{histogram.pdf}
 \caption{The absorption features appearing imprinted on the 0870870501 spectrum. The plot definition and colours are as per Fig.~\ref{fig:xmm_spectra}.}
 \label{fig:xmm0870870501_abs}
\end{figure}

As we mentioned above, the spectra appear to have some structured residuals to the best-fitting continuum model in {\it XMM-Newton} observation 0870870501, most notably at $\sim$1 keV and  $\sim$1.7 keV. We investigated this further by adding multiplicative gaussian absorption line components ({\sc gabs} in {\sc xspec}) to the model.  Interesting, the additional components do help to improve the fitting; the left panel of Fig.~\ref{fig:xmm0870870501_abs} shows the new best-fitting model together with the fitting residuals in which the best-fitting absorption line energies are at $0.99_{-0.04}^{+0.05}$ keV and $1.68\pm0.06$ keV.  The addition of both components improves the fit, with the absorption component at 0.99 keV improving the fitting statistic by $\Delta\chi^{2}$~=~$5.63$,\footnote{Here, we define $\Delta\chi^{2}$ as the difference between the $\chi^{2}$ of the models with and without the absorption components.} while that of the component at 1.68 keV changes the fitting statistic by $\Delta\chi^{2}$~=~$17.38$, for three fewer degrees of freedom in each case. Evidently, the putative absorption feature at 1.68 keV is relatively more significant than that at 0.99 keV. We also note that these residuals cannot be explained by either the complex, unresolved emission lines of a collisionally-ionised diffuse gas ({\sc apec} model in {\sc xspec}) or a warm absorber, as calculated by the {\sc XSTAR} model \citep{Kallman2001}.\footnote{\url{https://heasarc.gsfc.nasa.gov/docs/software/xstar/xstar.html}}

Our next step was to attempt to place better constraints on the significance of the putative absorption features. We used the {\sc xspec} command {\sc fakeit} to simulate pn, MOS1 and MOS2 spectra using the same observational response files as used for the real data, the best-fitting two component model from the 0870870501 data shown in Table~\ref{tab:two_component_fitting} (i.e., the two component model without the additional absorption line features), and an exposure time for each instrument exactly matching the real data (see Table~\ref{tab:obervations}). In addition, realistic statistical fluctuations were also added to the simulated spectra by {\sc fakeit}.  In total 60,000 sets of simulated spectra were created. Each set of fake observational spectra was then analysed by the same models used to analyse the real data, i.e. (i) two thermal components and (ii) two thermal components with an additional absorption component model; note that for the latter model, the parameters of the absorption component, including the line energy, were allowed to be free parameters during the fitting, similarly to the method that was used to analyse the real data, and the improvements in $\chi^2$ from the additional component were recorded.  Hence, we are able to determine the probability of detecting features that are simply statistical fluctuations in the simulated data, rather than real characteristics of the source. We found that the probability of obtaining values of $\Delta\chi^{2}$ $>$ 5.63 and $>$17.38 from statistical fluctuations alone are 16.28 per cent and 0.53 per cent respectively, corresponding to significance levels of $\sim 1.1\sigma$ and $\sim 2.6\sigma$.  Hence, there is no strong evidence that either feature is real, although the feature at 1.68 keV is at least marginally significant.  We will discuss the believability of this putative absorption feature and its possible physical origin further in Section~\ref{sec:Discussion}.

%%%%%%%%%%%%%%%%%%%%%%%%%%%%%%%%%%%%%%%%%%%%%%%%%%
%%%%%%%%%%%%%%%%%%%%%%%%%%%%%%%%%%%%%%%%%%%%%%%%%%

\subsubsection{Broad band spectral model} \label{sec:Broad band spectral model}

Near-simultaneous observations of the galaxy NGC 2403 were obtained with {\it XMM-Newton} and {\it NuSTAR} near to the end of October 2020 and April 2021.  NGC 2403 XMM4 lies in the field of view of both sets of detectors, so that they provide us with the opportunity to study the broadband spectra of this source. However, there is an offset of a few days between the start of observations by the two telescopes in both epochs. Given this, we grouped together two sets of data that had an offset of $\sim$1-2 days, which we hereafter refer as epoch 1 and epoch 2 data (see Table ~\ref{tab:obervations}). For the spectra in epoch 2 that have more than one observation per observatory, we combined all spectra and response files obtained from the same detector, using the {\sc sas} task {\sc epicspeccombine}  and the {\sc ftools} task {\sc addspec} for  {\it XMM-Newton}  and {\it NuSTAR} data, respectively. Throughout the analysis in this section, {\it XMM-Newton} and {\it NuSTAR} spectra from each epoch were modelled simultaneously within the full energy range of 0.3-30\,keV.

\begin{figure*}
 \includegraphics[width=9.2cm]{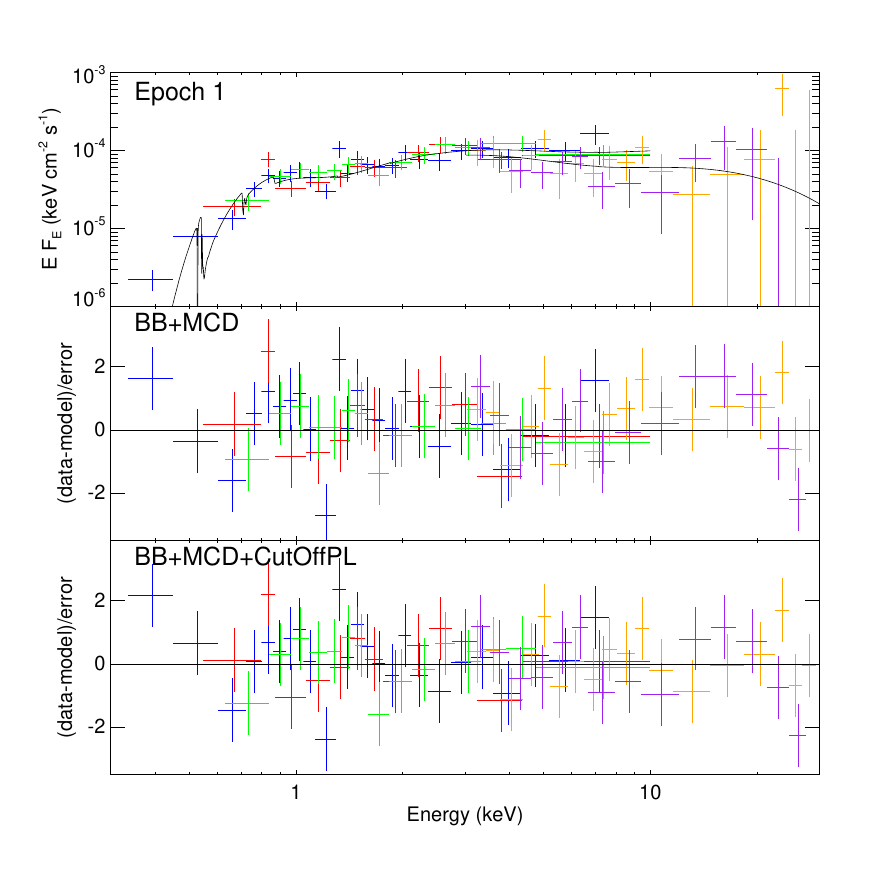}~\includegraphics[width=9.2cm]{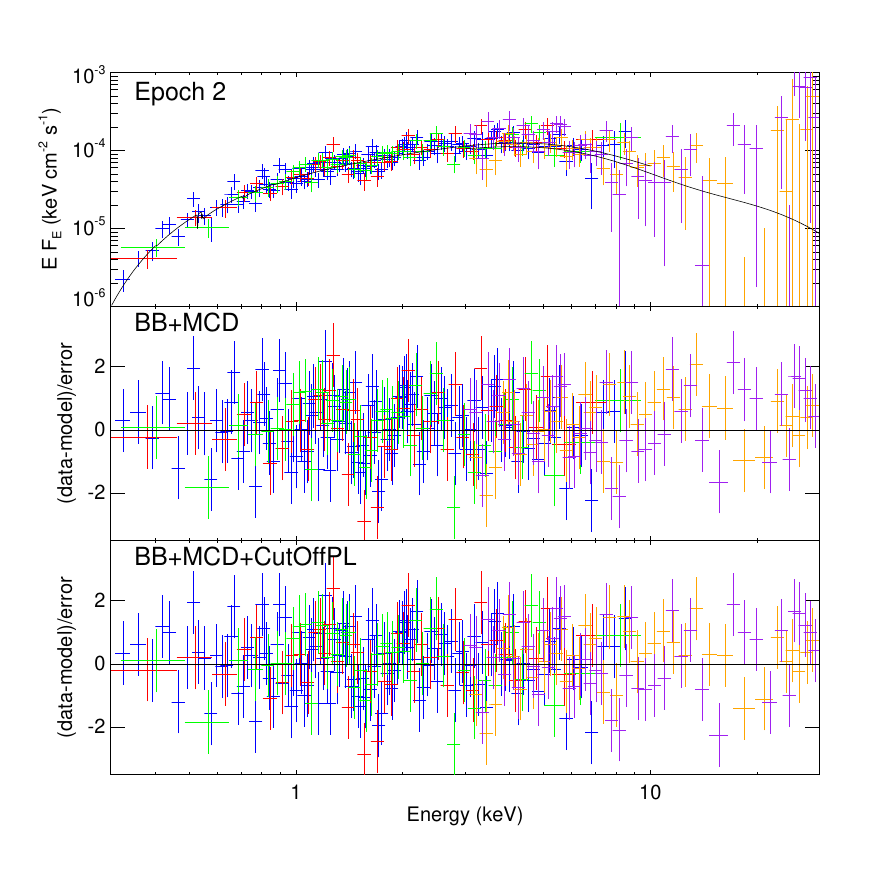}
 \caption{The broadband -- i.e., {\it XMM-Newton} $+$ {\it NuSTAR} -- spectra of NGC 2403 XMM4 fitted by the three component model (top panel) for observations of epoch 1 (left) and epoch 2 (right). The middle and bottom panels compare the residuals obtained from the best-fitting two and three component models. The orange and purple points present the {\it NuSTAR} FPMA and FPMB data, respectively, while the other colour codes and the plot definition are as per Fig.~\ref{fig:xmm_spectra}.}
 \label{fig:broadband_spectra}
\end{figure*}

We began by fitting an absorbed, two thermal component model -- exactly the same model as was used in Section~\ref{sec:Double component model} -- to examine whether a two component model is sufficient to explain the broadband spectra of the source, especially at the high energy end of the spectra.  The best-fitting result is shown in Table~\ref{tab:three_component_fitting}, along with the spectra and residual plots in Fig.~\ref{fig:broadband_spectra}. The reduced $\chi^{2}$ values in both epochs are $\sim$1, implying that the model is a reasonable description of the data. However, given both previous results for ULXs that show a hard excess above 10 keV \citep{Walton2018a}, and that, in particular, epoch 1 (Fig.~\ref{fig:broadband_spectra} middle panel) shows positive residuals at $\sim$10 -- 20 keV in which the model seems to underestimate the data, this convinced us to investigate further by adding a third component to the model -- an exponential cutoff power-law ({\sc cutoffpl} in {\sc xspec}) -- in an attempt to explain the spectra at high energies.  The addition of this component led to a $\Delta\chi^{2}$ of $\sim$10 for epoch 1, and $\sim$3 for epoch 2, for three fewer degrees of freedom.  This suggests that the additional component marginally improves the fit to the epoch 1 spectra, while there is little improvement in that of the epoch 2 spectra.  Thus, to examine whether the cut-off power-law component helps to improve the fit significantly, the {\sc xspec} statistical tool {\sc ftest}, i.e.,  the F-statistic probability of the fit improving given the $\Delta\chi^{2}$ and the change in degrees of freedom before and after adding the extra component, was applied as an first order approximation\footnote{For a discussion of the limitations on the statistical provenance of this technique, see \cite{Protassov2002}}.  Note that the {\sc xspec ftest} always calculates the probability in term of its {\it p}-value; here, the values of 1 - {\it p} are reported to better clarify the meaning of the probability.  The result is shown in column 8 of Table~\ref{tab:three_component_fitting}.  This does suggest that the epoch 1 spectra are likely to require the additional exponential power-law component to explain their high energy tail (with a probability of $\sim$98 per cent, corresponding to a confidence level of $\sim$2.3$\sigma$) while this component appears not to be needed for the epoch 2 spectra. We will discuss this point further in the discussion section.

%%%%%%%%%%%%%%%%%%%%%%%%%%%%%%%%%%%%%%%%%%%%%%%%%%
%%%%%%%%%%%%%%%%%%%%%%%%%%%%%%%%%%%%%%%%%%%%%%%%%%

%%%%%%%%%%%%%%%%%%%%%%%%%%%%%%%%%%%%%%%%%%%%%%%%%%

    \begin{table*}
      \centering
      \caption{The best fitting results from the broadband spectra modelled by two and three component models}\label{tab:three_component_fitting}
      \smallskip
      \begin{threeparttable}
          \begin{tabular}{lcccccccc}
          \hline
             Spectrum & $N_{\rm H}$ & $BB_{\rm Temp.}$ & $MCD_{\rm Temp.}$ & $\Gamma_{\rm cutoff}$$^{a}$ & $E_{\rm cutoff}$$^{b}$ & $\chi^{2}$/d.o.f. & Prob.$^{c}$  & $L_{\rm X}$ \\
             & ($\times$ 10$^{22}$ cm$^{-2}$)& (keV) &(keV)  & & (keV) &  & &($\times$ 10$^{39}$ erg s$^{-1}$)  \\
             \hline
             
\multirow{2}{*}{Epoch 1} & $<0.63$ & $0.17_{-0.05}^{+0.19}$ & $1.90_{-0.30}^{+0.46}$ &-- & -- & 80.15/76  & \multirow{2}{*}{0.98} &  \multirow{2}{*}{$0.68_{-0.07}^{+0.08}$} \\ 
				     & $0.70^{+0.62}_{-0.50}$ & $0.11_{-0.03}^{+0.06}$ & $1.02^{*}$ & $0.31_{-3.12}^{+1.96}$ & 6.90$^{*}$ & 70.36/73 & \\
	    \hline
\multirow{2}{*}{Epoch 2}& $0.04_{-0.03}^{+0.05}$ & $0.30_{-0.09}^{+0.12}$ & $1.87_{-0.13}^{+0.17}$ & -- & -- & 310.42/308 &  \multirow{2}{*}{0.63} &  \multirow{2}{*}{$0.80_{-0.03}^{+0.04}$} \\ 
				    & $0.05_{-0.02}^{+0.04}$ & $0.27_{-0.08}^{+0.13}$ & $1.62^{*}$ & 0.10$^{*}$ & 6.42$^{*}$ & 307.21/305   \\  
             
             \hline
         \end{tabular}
         \begin{tablenotes}
         \item \textit{Notes.} $^{a}$Photon index and $^{b}$ cut-off energy of the exponential cut-off power-law component. $^{c}$The probability indicating that the data would require the third cutoff power-law component to explain the spectra (see text for detail). $^{*}$No error bars reported as the parameter was fixed at its best-fit value to enable constraints to be placed on the confidence intervals of the other free parameters. The rest of the column definitions are as per Table~\ref{tab:two_component_fitting}.
         \end{tablenotes}
      \end{threeparttable}
    \end{table*}

%%%%%%%%%%%%%%%%%%%%%%%%%%%%%%%%%%%%%%%%%%%%%%%%%%

\subsection{Timing analysis}

\begin{figure}
\includegraphics[width=\columnwidth]{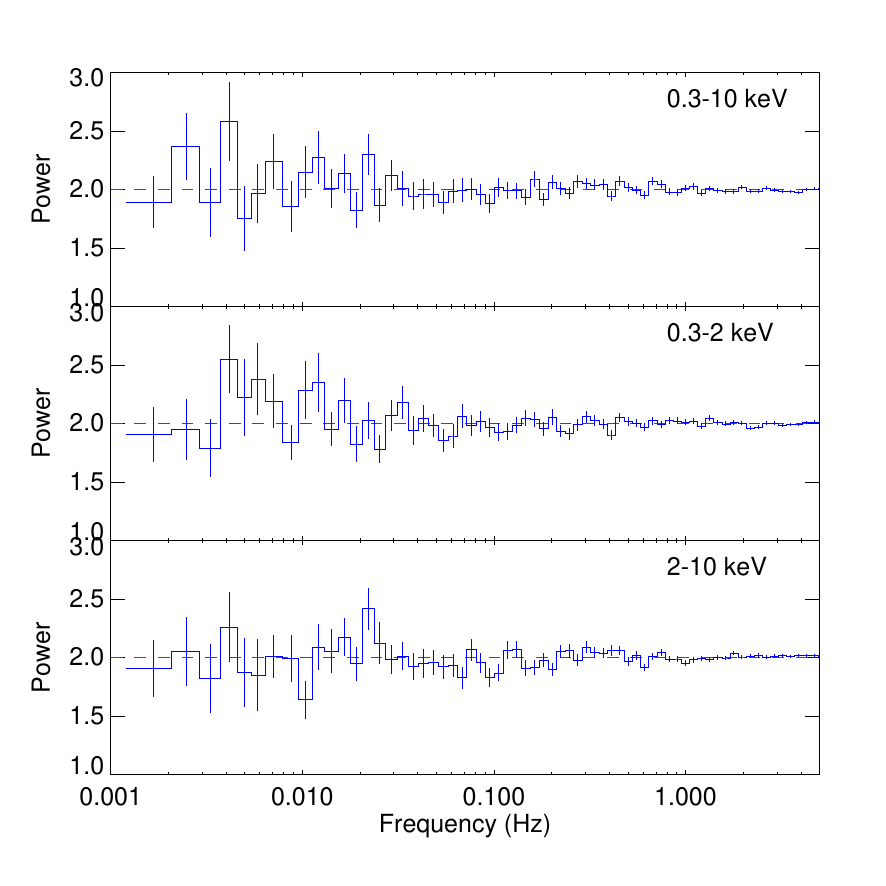}
 \caption{The PSDs of NGC 2403 XMM4 in the full (top), soft (middle) and hard (bottom) energy bands obtained from {\it XMM-Newton} observation 0164560901. The horizontal, red-dashed line represents the Poisson noise level.}
 \label{fig:psd}
\end{figure}

We present the long-term light curve of NGC 2043 XMM4 in Fig~\ref{fig:lightcurve}.  This demonstrates that it is a remarkably persistent source on timescales of days to years, with all observations over a $\sim 20$ year period having observed fluxes in the range $\sim 0.6 - 1.3 \times 10^{39}~\rm erg~s^{-1}$.  Here, we investigate whether this persistence is a characteristic down to short timescales, by analysing the timing information from the individual observations in Fourier space.  We start by searching for any significant variability in the observed frequency domain using the {\sc ftool} task {\sc powspec}, using the extracted light curves in the energy bands for each instrument defined in Section~\ref{sec:Observations and data reduction}, and applying the Leahy normalisation \citep{Leahy1983} in which the Poisson noise power is set equal to two.  Fig.~\ref{fig:psd} illustrates the power-spectral densities (PSDs) of {\it XMM-Newton} observation 0164560901 -- which is one of our highest quality timing datasets -- in the full, soft and hard energy bands. Unfortunately, there is no variability detected significantly above Poisson noise in any individual observation.  Although the analysis was also applied to analyse the concatenated light curves of all {\it XMM-Newton} and {\it Chandra} data, similarly, no variability was detected in these light curves either. We double-checked these results using an alternative software package, namely {\sc hendrics} version 7.0 \citep{Bachetti2018}\footnote{\url{https://hendrics.stingray.science/en/stable}} which is a set of command-line scripts based on the  {\sc stingray} Python library \citep{Huppenkothen2019a,Huppenkothen2019b,Bachetti2023}. No significant variability was found in any PSD created by this task either.

%%%%%%%%%%%%%%%%%%%%%%%%%%%%%%%%%%%%%%%%%%%%%%%%%%
%%%%%%%%%%%%%%%%%%%%%%%%%%%%%%%%%%%%%%%%%%%%%%%%%%

Although there is no periodical modulation hinted at by the calculated PSDs, we cannot immediately rule out NGC 2403 XMM4 as an X-ray pulsar.  PULXs do not often show strong pulsations in their PSDs as they tend to display strong period derivatives ($\dot{f}$), as a result primarily of orbital motion, that result in blurring of the pulsation power across multiple frequency bins (see e.g. \citealt{Sathyaprakash2019}).  Thus, we began searching for a pulsating signal in the frequency range of 0.1 -- 5 Hz using data from the only instruments with sufficient temporal resolution to investigate this regime, the {\it XMM-Newton} pn and {\it NuSTAR} FPMA and FPMB detectors.  We used the full energy band of the data (i.e., 0.1 -- 12 keV for pn data and 3 -- 79 keV for FPMA \& FPMB) and analysed it using the {\sc hendrics} task {\sc henaccelsearch} which utilises the accelerated search method detailed in \citet{Ransom2002}.  Subsequently, any detected candidates were further analysed and verified by the {\sc hendrics} epoch folding algorithm {\sc henzsearch} which applies the $Z^{2}$ statistic with $n$ harmonics (c.f. \citealt{Buccheri1983});  here $n$$=$3 (i.e., $Z^{2}_{3}$) was used.

\begin{figure*}
\includegraphics[width=\textwidth]{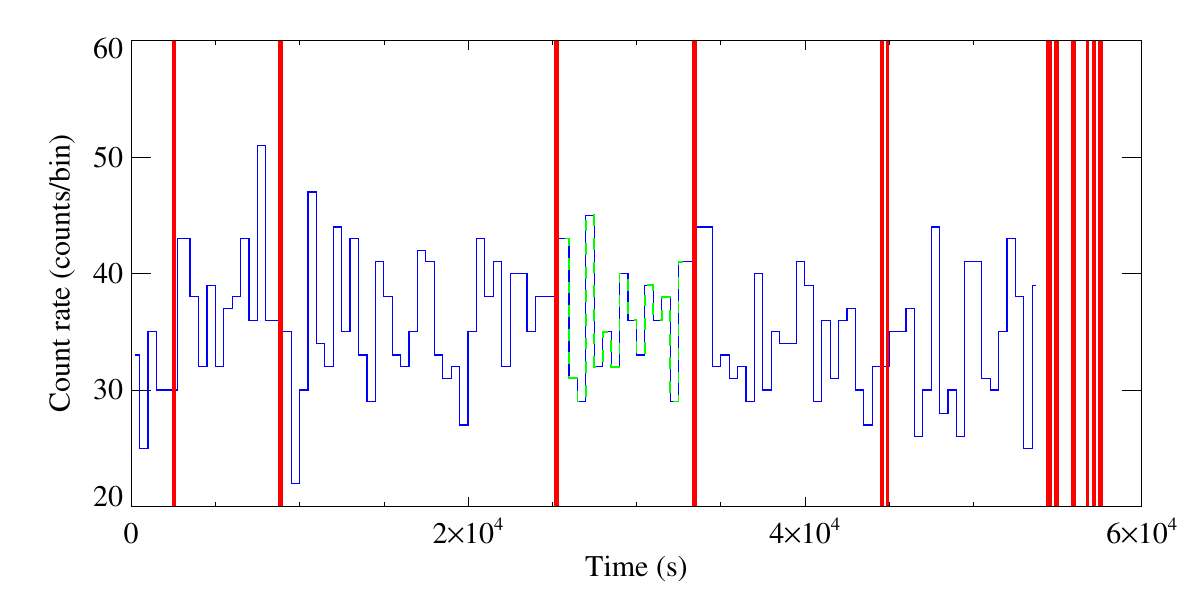}
 \caption{The light curve of NGC 2403 XMM4 with 500 s bin size obtained from {\it XMM-Newton} data Obs. ID 0164560901. The red vertical lines represent the boundaries of bad time intervals. The dashed-green line highlights the GTI in which the pulsating signal was detected.}
 \label{fig:pulse_lightcurve}
\end{figure*}

Using the above method, the search was applied to the whole length of each individual observation, but no significant pulsation could be verified in any searches. However, as illustrated in e.g. \citet{Sathyaprakash2019}, the source might not exhibit a persistent observable pulsating signal throughout the entire observational period and, indeed, the signal might not be detected in all observational data; instead, the pulsating power might only be strong enough to be detected during some discrete periods. To check this, we divided each observational dataset into a number of segments. Instead of dividing each observation into the segments with equal time length, we used the gap between GTIs -- i.e., observational periods with bad background flaring or no data -- as boundaries for segmenting each dataset.  This is to ensure that the gaps in the data are excluded from the calculations. Fig.~\ref{fig:pulse_lightcurve} illustrates how the observational data were divided into segments using this method; note that, while most segments are sufficiently long to be useful for the calculations, some segments are too short ($\sim$0.1--0.3 ks) and so were excluded from the analysis.

\begin{figure}
\includegraphics[width=\columnwidth]{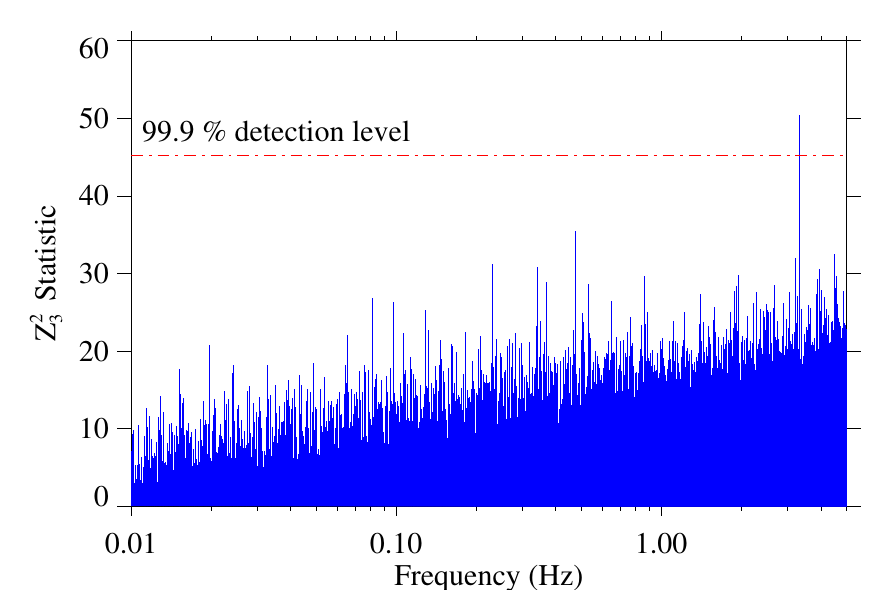}
\includegraphics[width=\columnwidth]{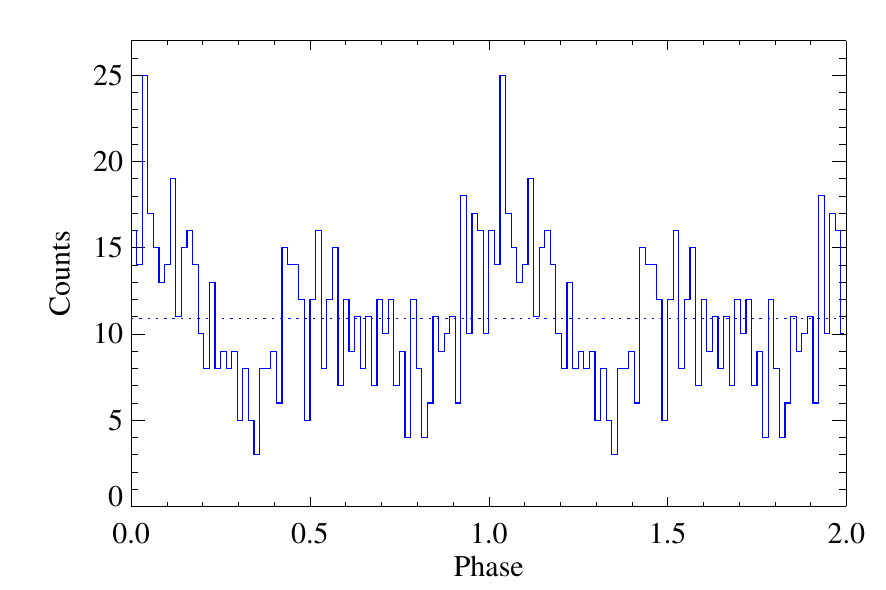}
 \caption{Top panel: the $Z^2_3$ statistic as a function of frequency, calculated for the segment of {\it XMM-Newton} observation 0164560901 in which a pulsation at $\sim$3.3 Hz is detected. Bottom panel: the folded pulse profile during the period that the pulsating signal is detected; the horizontal dotted line demarcates the mean count rate.}
 \label{fig:pulse_detect}
\end{figure}

\begin{figure*}
\includegraphics[width=15cm]{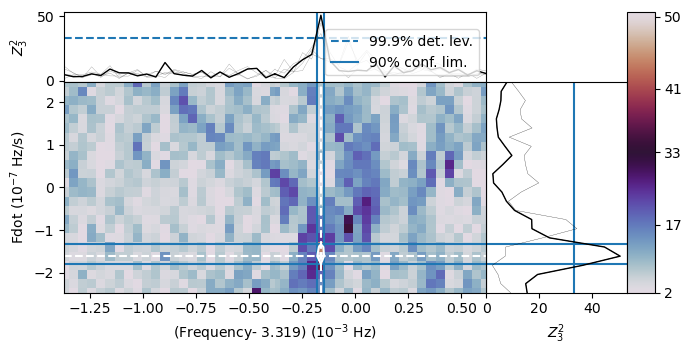}
 \caption{The $f$-$\dot{f}$ parameter space showing the values of $f$ and $\dot{f}$ determined for the segment of observation 0164560901 in which the source exhibited a pulsation.}
 \label{fig:f-fdot plot}
\end{figure*}

By dividing the observational data into segments and applying the search again, fascinatingly, there is one time segment in the {\it XMM-Newton} data of observation 0164560901 (observed in September 2004) which exhibited a detectable pulsating signal -- i.e. the fourth segment in Fig.~\ref{fig:pulse_lightcurve} -- at $\sim$ 3.32 Hz (i.e., $\sim$0.3 s period) with a $Z^{2}_{3}$ power of 50.6, corresponding to a statistical significance for the detection of $>$99.9 per cent (see Fig.~\ref{fig:pulse_detect}, top panel). The bottom panel of Fig.~\ref{fig:pulse_detect} presents the folded pulsed profile, in which the determined pulsed fraction is 38.1 $\pm$ 4.3 per cent, and the pulse itself appears rather discrete; it is not obviously sinusoidal in profile, as per most PULXs (e.g. \citealt{Bachetti2014}). In addition, we also plot the $f$-$\dot{f}$ parameter space in Figure~\ref{fig:f-fdot plot} to demonstrate the peak location where the statistical significance of these two parameters are highest; the analysis constrains values of $f$ of $3.3188_{-2.26\times10^{-6}}^{+1.95\times10^{-6}}$~Hz and the $\dot{f}$ of -1.61$_{-0.0240}^{+0.0574}$~$\times10^{-7}$ Hz s$^{-1}$. Note that the range over which the dataset segment was searched was $f = 3$ -- 3.5 Hz with resolution of $\sim$ 3.95 $\times$ 10$^{-5}$  Hz, while that of $\dot{f}$ was $\sim$($-$2.46 -- $+$2.46) $\times$ 10$^{-7}$ Hz s$^{-1}$ with resolution of $\sim$2.14 $\times$ 10$^{-8}$  (i.e., 291,456 trials in total). We emphasise that there is no other significant pulsation detected in any other segment of this observation, or in any segments of the other observations, including those of the {\it NuSTAR} data. However, we do note that there were at least two lower significance detections with $Z^{2}_{3}$ powers of $\la$ 40, detected at frequencies of $\sim$3.2 Hz and $\sim$3.4 Hz in segments of the data from {\it XMM-Newton} observations 0870870501 and 0870870801.  Although it is very encouraging that we detect further signals in close proximity to the frequency of our one significant detection (see \citealt{Sathyaprakash2019} for the importance of such detections in confirming real, weak pulsations), their low statistical significance is insufficient in either case to confirm they constitute real pulsating signals.

%%%%%%%%%%%%%%%%%%%%%%%%%%%%%%%%%%%%%%%%%%%%%%%%%%
%%%%%%%%%%%%%%%%%%%%%%%%%%%%%%%%%%%%%%%%%%%%%%%%%%

%%%%%%%%%%%%%%%%%%%%%%%%%%%%%%%%%%%%%%%%%%%%%%%%%%
%%%%%%%%%%%%%%%%%%%%%%%%%%%%%%%%%%%%%%%%%%%%%%%%%%

%%%%%%%%%%%%%%%%%%%%%%%%%%%%%%%%%%%%%%%%%%%%%%%%%%
%%%%%%%%%%%%%%%%%%%%%%%%%%%%%%%%%%%%%%%%%%%%%%%%%%

\section{Discussion} \label{sec:Discussion}

In this work, we have studied observational data for the X-ray source NGC 2403 XMM4 obtained from the archives of multiple X-ray telescopes from the current generation of observatories.  Combining the variety of data we obtained provides a broadband energy range from soft to hard X-rays (i.e., $\sim$ 0.3 -- 30 keV), and temporal resolution ranging from as short as $\sim$ 5 Hz to long timescales of up to two decades.  This range of data permits us to conduct the variety of spectral and temporal analysis shown in the previous section.  Here, we discuss what they mean for the nature of this object.

\subsection{Should we regard this source as a ULX?} \label{is it a ULX}

Our interest in NGC 2403 XMM4 originates from its inclusion in several catalogues and/or studies of ULXs (see e.g. \citealt{Walton2011,Winter2006}).  However, given the 20 year light curve of the source we show in Fig.~\ref{fig:lightcurve}, it is obvious that, most of the time, the source luminosity lies below 10$^{39}$ erg s$^{-1}$, and hence is below the threshold luminosity normally used to define ULXs.  We therefore begin by posing a question: should we continue to regard this object as a ULX?

The only data points in Fig.~\ref{fig:lightcurve} consistent with a luminosity above the threshold for ULXs were obtained from {\it Swift} observations, and all of those may be consistent with luminosities below this, given the low number of counts per {\it Swift} observation and hence relatively high uncertainties.  Indeed, we cannot reject the null hypothesis that the source luminosity is persistently below 10$^{39}$ erg s$^{-1}$ at 3$\sigma$ confidence from this dataset.  Furthermore, there is a possibility that we have overestimated the distance to NGC 2403.  Our value of 4.2 Mpc -- which is adopted from the \citet{Walton2011} catalogue -- is actually at the higher end of the range of measurements for the distance to NGC 2403, as reported in the NASA/IPAC Extragalactic Database (NED)\footnote{\url{https://ned.ipac.caltech.edu}}, which typically range from $\sim3 - 4$ Mpc (e.g. \citealt{Vaucouleurs1981,Bottinelli1984,Olivares2010}). Thus, if the source distance is reduced by a factor of $3/4$, the luminosity would be reduced by a factor of $(3/4)^{2}$ $\sim$ 0.6. This would imply that the source luminosity lies well below the traditional ULX luminosity regime, at $\sim 5 \times 10^{38}$ erg s$^{-1}$.  Indeed, this is very likely why this source is included in the \citet{Walton2011} catalogue, but not in more recent ULX catalogues (e.g. \citealt{Earnshaw2019,Walton2022}) despite being detected in multiple serendipitous surveys.

It might however be argued that 'ultraluminous' is a better description of a behaviour of an object, than of the nature of the object itself; indeed it has long been suggested that ULXs represent a particular behaviour of stellar-mass compact objects (e.g. the 'ultraluminous state' of \citealt{Gladstone2009}).  There is now evidence of many objects catalogued as ULXs occasionally (or in some cases frequently) dropping to luminosities below the ULX regime, for example some of the Eddington threshold objects studied by \cite{Earnshaw2017}, or the sample of highly variable ULXs presented by \cite{Song2020} that notably highlights that several of the known PULXs do this.  Hence, it is still plausible that NGC 2403 XMM4 is a ULX, at least some of the time, although the evidence based solely on its luminosity is somewhat tenuous.

\subsection{Does NGC 2403 XMM4 behave like a ULX?}

The NGC 2403 XMM4 spectra can be modelled by two-component models composed of two thermal components, consistent with ULXs (e.g. \citealt{Stobbart2006,Koliopanos2017}), as is illustrated by Fig.~\ref{fig:xmm_spectra} and Fig.~\ref{fig:L-T_correlation}.  The spectra of this source appear disc-like at higher luminosities, but display two distinct components at lower luminosity.  By comparison to the spectral regimes proposed by \citet{Sutton2013} for ULXs, the former and latter spectral shapes can be classified as broadened disc and hard ultraluminous, respectively. In their study, \citet{Sutton2013} showed that the dominant spectra of ULXs (as a population) changed from broadened disc below $\sim 3 \times 10^{39} \rm ~erg~s^{-1}$ to hard (or soft) ultraluminous as the luminosity increases, which they suggested might be an evolution from sources accreting at around the Eddington rate to overtly super-Eddington accretion flows.  However, subsequent work demonstrated that the spectra of BHBs in the thermal-dominant state, typically accreting matter at $\sim$10 -- 30 per cent of their Eddington rate (and so having luminosity $\sim$ 10$^{38}$ erg s$^{-1}$), exhibit a broadened-disc like spectral shape, similar to that of the low-luminosity ULX spectra, in the 0.3-10\,keV band that ULXs are observed in \citep{Sutton2017}.  Hence, this spectrum could suggest that the source is a BH binary accreting at moderately high rates, but still in the sub-Eddington regime. However, studies of individual ULXs (see e.g. \citealt{Luangtip2016}) have demonstrated that these objects evolve from a distinct two-component (hard ultraluminous) spectrum at lower luminosities, to a single broad disc-like component (broadened disc) at the highest luminosities.  We cannot rule out that that is happening for NGC 2403 XMM4, despite it much lower luminosity.

A further hint that NGC 2403 XMM4 is accreting at super-Eddington rates comes from broadband spectroscopy.  The presence of a steep high energy tail, albeit only at moderate statistical significance ($\sim 2.3\sigma$) in one of two observational epochs, is another similarity to ULX spectra.  Such a component is a general property of ULXs and PULXs, in addition to the two thermal components below 10 keV (e.g. \citealt{Mukherjee2015,Walton2017,Koliopanos2019,Walton2020}).  This high energy tail can be explained in the context of Compton up-scattering in an X-ray corona (a similar mechanism to that found in sub-Eddington BH binaries) or a super-Eddington accretion column onto a magnetised NS \citep{Walton2020}; the former scenario could be possible for both BH and non-magnetic NS accretors whilst the latter requires that the accretor must be a magnetised NS.

Moreover, after careful removal of the outlier datasets, we found a correlation between the flux and temperature of the soft component of $L$ $\propto$ $T^{1.49\pm0.85}$.  This result is broadly consistent with those found in previous ULX studies, including for PULXs (e.g. \citealt{Pintore2012,Walton2013,Walton2020,Robba2021,Barra2022}). Although many of these these studies detect a putative $L-T$ correlation for both soft and hard components, in our case a correlation is only found for the soft component, which may be related to the small number of data points and their relatively low spread of values in the warm component.  Interestingly, the temperature power index of $\sim$1.5 is consistent with one report of a similar relationship for the PULX NGC 1313 X-2, in which an index of $1.2\pm0.3$ was reported by \citet{Pintore2012} despite leaving $N_{\rm H}$ free to vary; however a subsequent analysis of a larger dataset for this source found a very different $L \propto T^{-3.9\pm1.0}$ relationship \citep{Barra2022}.  The relationship we derive for NGC 2403 XMM4 is consistent within errors with the $L$ $\propto$ $T^{2}$ prediction for advection-dominated (slim) disc models \citep{Watarai2000,Watarai2001}, that may form at the centre of super-Eddington accretion flows \citep{Poutanen2007}.  It is only very marginally consistent with a standard disc model where $L$ $\propto$ $T^{4}$ \citep{Shakura1973}, at the $\sim 2.95 \sigma$ level.  This behaviour therefore also points to the possibility that the accretion in NGC 2043 XMM4 is super-Eddington.

\subsection{Is this a super-Eddington neutron star?} \label{sec:Is this a super-Edd NS?}

Although NGC 2403 XMM4 appears to spend little time (if any) above the $10^{39} \rm ~erg~s^{-1}$ threshold for a ULX, its spectral behaviour does appear distinctly ULX-like, which may imply that it is accreting at super-Eddington rates.  If so, we may estimate its compact object mass simply by using equation 1 of ~\citeauthor{Feng2011} (\citeyear{Feng2011}; see also \citealt{Frannk2002}),  assuming that the time averaged luminosity of NGC 2403 XMM4 is $\sim$5 $\times$ 10$^{38}$ erg s$^{-1}$ and the emission is isotropic.  This luminosity then corresponds to a compact object mass $<$3.8M$_{\rm \odot}$ for an object accreting matter at super-Eddington rates, or even lower if the X-ray emission is beamed.  This implies that the compact object is a NS; for a theoretical mass range for NSs of 1.2 -- 2.9M$_{\rm \odot}$ \citep{Kalogera1996,Suwa2018}, the typical luminosity of $\sim$5 $\times$ 10$^{38}$ erg s$^{-1}$ corresponds to accretion at between 1.3 -- 3.2 times Eddington; for peak luminosities up to $\sim 10^{39} \rm ~erg~s^{-1}$, this can go as high as 6.4 times the Eddington rate.

There are two key pieces of evidence that we have not discussed yet that support this diagnosis.  The first, most obviously, is the putative detection of pulsations during a short segment of the {\it XMM-Newton} observation 0164560901.  If real, this is clear evidence for a NS that is accreting at super-Eddington rates.  Our detected signal was at a frequency of 3.32 Hz (corresponding to a period of 0.30 s) with frequency derivative of -1.61 $\times$ 10$^{-7}$ Hz s$^{-1}$ and pulse fraction of $\sim$38 per cent.  This was detected with a statistical value of $Z^{2}_{3}$ $\sim$ 51, corresponding to a detection significance level of $>$ 99.99 per cent.  

However, its detection in only one segment of one out of many observations may mean its pulsations are transient.  Similar behaviour has been seen in other PULXs (and, indeed, PULX candidates - see \citealt{Bachetti2014,Sathyaprakash2019,Quintin2021}), although the cause of this behaviour remains a puzzle.  One possible clue is that the pulse fraction of the PULX M82 X-2 is proportional to its intensity (\citealt{Bachetti2020}; see also \citealt{Lewin1988}). Interestingly, the data in which we detected the pulsation is the highest flux {\it XMM-Newton} observation of the source (cf. Table~\ref{tab:two_component_fitting}), which might be a factor in the pulse being detected; however it does not explain why this only happens in a short segment of the data, given that the source flux did not change significantly during the observation (see e.g. Fig.~\ref{fig:pulse_lightcurve}). One possibility to explain this relates to the scenario proposed by \citet{King2020}, where they suggest that the pulsations of PULXs are only detectable if the NS spin axis and the accretion disc beaming axes are strongly misaligned; otherwise, the pulse fraction would be small (or tend to zero), and so pulsations would not be detectable. Thus, if the angle between the the NS spin and the accretion disc beaming axes could dynamically change with time, perhaps due to spin axis precession on hour timescales similar to that detected in some NSs (e.g. \citealt{Heyl2002,Makishima2014}), a pulse on/off phase might be seen in the data. Alternatively, variations in the obscuration towards the pulsing regions caused by the inner disc structure and/or an outflowing wind along the line of sight might cause the pulsations to become visible for a short period of time (e.g. \citealt{Kosec2018,Walton2018a,Barra2022}).  

The other plausible explanation is that the pulsation is not real - indeed, by slicing the multiple datasets into many more datasets, and then searching an extensive region of $f - \dot{f}$ parameter space, we should perhaps expect to see some false positive signals at relatively high significance. A full analysis of the statistics of false detections is beyond the scope of this paper; however we can at least examine whether, given a candidate signal close to 3.3 Hz, how likely a purely stochastic false signal would be in our data.  To do this, we simulated 10,000 fake datasets with the same properties -- i.e. mean count rate and exposure time -- as the real data segment that the pulsation was detected in, using the {\sc hendrics} script {\sc henfake}; the fake data were then searched for a pulsating signal using the same method applied to the real data. We find a probability that random fluctuations would produce a false pulsation of frequency $\sim$3.3 $\pm$ 0.25 Hz with a power of $\geqslant$ 50.6 is 0.001, corresponding to a confidence level of 3.0$\sigma$ that the detected pulsation is real. The other key to the believability of such signals is the detection of another signal at a very similar frequency in another segment/observation (cf. \citealt{Sathyaprakash2019}).  This is hinted at in two other segments, but neither signal is statistically robust.  We must therefore conclude that this signal marks NGC 2043 XMM4 as a candidate pulsating object, but this remains to be confirmed.  It is, however, another hint that this object may host a NS.

Assuming that the detected pulsation is genuine, compared to other super-Eddington NSs with pulsations -- the PULXs -- this has the fastest pulsations (marginally faster than, but still comparable with, the $\sim 0.4$ s period of NGC 7793 P13) with a relatively high pulse fraction at $\sim 38$ per cent. Some theoretical studies have explained the observed properties of PULXs in the context of an accreting magnetar (see e.g. \citealt{DallOsso2015,Eksi2015,Mushtukov2015}, but see also \citealt{King2019}), in which it would be expected to spin up as material from the donor is accreted over time. This behaviour has also been seen in the long term ($\sim$ 4 years) monitoring campaign of NGC 7793 P13 \citep{Furst2021}.  In fact, calculations by \citet{Meng2022} predicted that 10$^{4}$ years ago the source was a highly magnetised NS with an initial magnetic field strength and pulse period of 10$^{14}$ G and $\sim$ 10 s, respectively -- similar to those of the other confirmed PULXs; as the source evolves with continuous accretion, the magnetic field decays while its spin becomes faster. This would imply that most detected PULXs are relatively young, compared to NGC 7793 P13 as well as (potentially) NGC 2403 XMM4, which exhibit relatively fast spin periods, but lower magnetic fields (see fig. 2 of \citealt{Meng2022}).  However, NGC 1313 X-2 would be very anomalous in this context given that it is surrounded by a bubble nebula that implies an age in excess of $10^5$ years for that PULX \citep{Sathyaprakash2019}.

The other potentially important piece of evidence is the possible detection of absorption features in the spectrum from {\it XMM-Newton} observation 0870870501, in particular the feature at 1.68 keV with moderate ($2.6\sigma$) significance (for this discussion, we will disregard the minimally significant feature at 0.99 keV).  We first caution that this line energy is close to known instrumental features of {\it XMM-Newton}\footnote{\url{https://xmmweb.esac.esa.int/docs/documents/CAL-TN-0018.pdf}}, however these features should be properly addressed during the analysis.  Indeed, it is clear that we did not see this feature in the other {\it XMM-Newton} observations, implying that the feature is intrinsic to the source during that one observation (if it is a real feature).  

It is interesting to speculate whether the absorption feature could be a CRSF produced by the strong magnetic field of a NS, similar to those reported by \citet{Brightman2018,Walton2018b}.  If it were, it would imply that NGC 2403 XMM4 is a magnetised NS, and would implicitly support the detected pulsation as a genuine feature, since it could be generated by a spinning magnetised NS. If we assume that the line absorption is a result of the cyclotron resonance scattering mechanism, we can determine the magnetic field close to the NS surface.  Assuming electron scattering, $B_{\rm 12}$~(G)~=~$E_{\rm CRSF}$~(keV)~$\times$~($1+z_{\rm grav}$)~/~11.57, where $B_{\rm 12}$ is the magnetic field in unit of 10$^{12}$ G; $E_{\rm CRSF}$ is the line energy in keV; and $z_{\rm grav}$ is gravitational redshift which is assumed to be $\sim$0.25 \citep{Brightman2018,Walton2018b}. This predicts a NS surface magnetic field of 1.82 $\times$ 10$^{11}$ G. However, we note that such a strong magnetic field could also actually beam the emitted X-ray photons, resulting in anisotropic emission (see e.g. \citealt{Mushtukov2022}). Assuming a nearly face-on, 2.9M$_{\rm \odot}$ NS accreting matter at $\sim$Eddington rate, a beaming factor of $\sim$1.3 (or higher for a typical less massive and/or sub-Eddington accreting NS) would be required for the source to reach an average luminosity of 5 $\times$ 10$^{38}$ erg s$^{-1}$. Thus, if the CRSF scenario is correct, we cannot completely rule out that this source is a $\sim$Eddington or sub-Eddington NS accretor.

However, an alternative interpretation of the 1.68 keV feature (which may also explain the very marginal feature at 0.99 keV) could be that the feature is produced by a combination of atomic features, in both absorption and emission, with the absorption features (at least) resulting from material in the fast ($v\sim 0.2c$) outflowing wind generated by super-Eddington accretion sources, and so imprinted on its observed X-ray spectrum (see e.g. \citealt{Pinto2016}).  At the spectral resolution of the {\it XMM-Newton} EPIC detectors individual features are not often resolved, but broader features resulting from a combination of discrete lines are seen in ULX spectra \citep{Middleton2014,Middleton2015b}.  In fact, referring to Fig.~3 of \citet{Pinto2016}, it can be seen in the {\it XMM-Newton} RGS spectrum of the ULX NGC 1313 X-1 that an absorption line (with $> 4\sigma$ detection significance) has been found at a wavelength of $\sim$7.5\AA~(1.65 keV)\footnote{A significant absorption feature with this energy is also detected in the RGS spectrum of the ULX NGC 5408 X-1 (see Fig.~6 of the same paper).} which can be identified as Ne X$\beta$ with a blueshift of $\sim$0.2$c$.  Interestingly, this absorption line is consistent within errors with the detected feature in NGC 2403 XMM4 at $1.68\pm0.06$ keV.  The feature therefore presents possible evidence for an outflowing wind similar in nature to those seen in ULXs; if, similar to ULXs, it is driven from the accretion flow by the extreme radiation pressure of a super-Eddington flow, then this again supports the notion of the accretor being a NS.  Indeed, it is notable that both scenarios for the absorption feature outlined here support the presence of a NS accreting at super-Eddington rates.

\section{Conclusion} \label{sec:Conclusion}

Our interest in NGC 2403 XMM4 was piqued by its inclusion in some older ULX catalogues and studies; however in this work we demonstrate that a classification as a ULX based purely on luminosity is questionable for this source, given that it rarely (if ever) reaches the $10^{39} \rm ~erg~s^{-1}$ threshold.  However, despite not making this threshold, we show that the object displays many behaviours and characteristics that are consistent with super-Eddington accretion: a two-thermal-component spectrum in the 0.3-10\,keV regime with a possible steep high energy tail above 10 keV; a soft component whose luminosity and temperature vary together in a manner consistent with an advection-dominated (slim) disc, at least when the spectra are in a similar disc-like state; one observation showing a putative absorption feature, which could be interpreted as a CRSF or absorption in an outflowing wind; and a possible transient pulsation at a similar frequency to PULXs.  Although individually each of these results might be questionable -- continuum spectra tend to be degenerate to multiple models, the absorption feature is only moderately significant, and the pulsation is only detected in one epoch and so requires a second detection for confirmation it is real -- together they all point in one direction, which is that NGC 2403 XMM4 is a NS that is accreting at super-Eddington rates.  If so, it is the third such source detected in NGC 2403.

\section*{Acknowledgements}

We thank the anonymous referee for their constructive comments, that have helped improve this paper. WL acknowledges the support of the Centre for Extragalactic Astronomy (CEA) at Durham University for hosting him during the preparation of this paper. TPR acknowledges support from STFC as part of the consolidated grant award ST/X001075/1.  Part of the data in this work are based on observations obtained with {\it XMM–Newton}, an ESA science mission with instruments and contributions directly funded by ESA Member States and NASA. This research has made use of data obtained from the {\it Chandra} Data Archive  and software provided by the {\it Chandra} X-ray Center (CXC) in the application package CIAO. We also acknowledge the use of public data from the {\it Swift} and {\it NuSTAR} data archives.

%%%%%%%%%%%%%%%%%%%%%%%%%%%%%%%%%%%%%%%%%%%%%%%%%%
\section*{Data Availability}

The data underlying this article were accessed from the {\it XMM-Newton}, {\it Chandra}, {\it NuSTAR} and {\it Swift} observatory archives (URLs given in Section~\ref{sec:Observations and data reduction}). The derived data generated in this research will be shared on reasonable request to the corresponding author.



\bibliographystyle{mnras}
\bibliography{references} % if your bibtex file is called example.bib

%%%%%%%%%%%%%%%%%%%%%%%%%%%%%%%%%%%%%%%%%%%%%%%%%%

%%%%%%%%%%%%%%%%% APPENDICES %%%%%%%%%%%%%%%%%%%%%

\appendix

\section{{\it Swift} observations analysed in this study} \label{sec:swift_obs}

The data from {{\it Swift} XRT observations analysed in this work can be obtained at \url{url}.

%%%%%%%%%%%%%%%%%%%%%%%%%%%%%%%%%%%%%%%%%%%%%%%%%%

    \begin{table*}
      \centering
      \caption{{\it Swift} observations of NGC 2403 XMM4}\label{tab:swift_obs}
      \smallskip
      \begin{threeparttable}
          \begin{tabular}{lcccc}
          \hline
Observation ID & Observed date &  Exposure time   & XRT Count rate & Flux   \\
             &                  &       (ks)          &    ($\times$ 10$^{-3}$ counts s$^{-1}$) &      ($\times$ 10$^{-13}$ erg s$^{-1}$ cm$^{-2}$)  \\
             \hline
         
00031418001 & 2009-06-11  & 4.79 & 8.64 & $5.38_{-1.64}^{+2.05}$ \\ 
00031418002 & 2009-09-15  & 5.28 & 6.29 & $3.21_{-1.03}^{+1.25}$ \\ 
00031418003 & 2009-12-15  & 4.90 & 5.17 & $2.85_{-0.92}^{+1.15}$ \\ 
00035870002 & 2006-10-09  & 2.29 & 6.71 & $4.59_{-1.77}^{+2.33}$ \\ 
00035870003 & 2006-10-15  & 5.69 & 8.91 & $5.35_{-1.22}^{+1.41}$ \\ 
00035870004 & 2007-04-06  & 2.34 & 6.31 & $3.50_{-1.46}^{+1.94}$ \\ 
00036563001 & 2007-12-02  & 6.69 & 8.26 & $5.01_{-1.15}^{+1.34}$ \\ 
00082262001 & 2013-11-29  & 5.27 & 4.92 & $3.13_{-1.00}^{+1.25}$ \\ 
00082262002 & 2013-12-01  & 8.18 & 4.62 & $3.13_{-0.79}^{+0.93}$ \\ 
00082262003 & 2013-12-03  & 1.25 & 3.37 & $<$4.38 \\ 
00082262004 & 2013-12-04  & 2.96 & 5.99 & $3.95_{-1.71}^{+2.39}$ \\ 
00082262005 & 2013-12-06  & 3.12 & 6.87 & $4.39_{-1.63}^{+2.15}$ \\ 
00089149001 & 2021-04-27  & 1.71 & 7.45 & $4.81_{-2.10}^{+2.71}$ \\ 
00089149002 & 2021-04-29  & 1.80 & 6.18 & $3.99_{-1.91}^{+2.71}$ \\ 
00095666001 & 2020-04-14  & 1.66 & 7.58 & $4.14_{-1.80}^{+2.54}$ \\ 
00095666002 & 2020-05-17  & 0.48 & 5.98 & $<$33.99 \\ 
00095666003 & 2020-05-26  & 0.64 & 3.41 & $<$33.71 \\ 
00095666004 & 2020-05-27  & 0.96 & 7.51 & $<$7.38 \\ 
00095666006 & 2020-10-06  & 1.50 & 5.37 & $<$5.51 \\ 
00095666007 & 2020-10-10  & 0.96 & 5.04 & $<$8.06 \\ 
00095666008 & 2020-11-10  & 2.29 & 3.22 & $<$5.19 \\ 
00095666009 & 2021-01-19  & 2.57 & 4.10 & $2.83_{-1.29}^{+1.74}$ \\ 
00095666010 & 2021-02-23  & 2.19 & 9.78 & $5.94_{-2.51}^{+3.45}$ \\ 
03109691001 & 2019-08-07  & 1.10 & 4.76 & $<$10.68 \\ 
03109691002 & 2019-08-09  & 0.75 & 3.15 & $<$4.55 \\ 
03109691003 & 2019-09-25  & 1.29 & 6.22 & $<$8.51 \\ 
03109691004 & 2019-11-20  & 0.76 & 3.30 & $<$9.30 \\           
            
           \hline
         \end{tabular}
      \end{threeparttable}
    \end{table*}

%%%%%%%%%%%%%%%%%%%%%%%%%%%%%%%%%%%%%%%%%%%%%%%%%%

%%%%%%%%%%%%%%%%%%%%%%%%%%%%%%%%%%%%%%%%%%%%%%%%%%

% Don't change these lines
\bsp	% typesetting comment
\label{lastpage}
\end{document}